\documentclass{jpp}
\usepackage{graphicx}

\usepackage{enumitem}
\usepackage{floatrow}
\usepackage[export]{adjustbox}

\usepackage{caption}
\usepackage{subcaption}
\usepackage{hyperref}
\hypersetup{
    pdftoolbar=true,        
    pdfmenubar=true,        
    pdffitwindow=false,     
    pdfstartview={FitH},    
    pdftitle={My title},    
    pdfauthor={Jeffersson Andrés Agudelo Rueda},     
    pdfsubject={Subject},   
    pdfcreator={Creator},   
    pdfproducer={Producer}, 
    pdfkeywords={keyword1, key2, key3}, 
    pdfnewwindow=true,      
    colorlinks=true,       
    linkcolor=black,          
    citecolor=blue,        
    filecolor=magenta,      
    urlcolor=cyan           
    pdfborder={0 0 0}
    hidelinks=true
}

\usepackage[utf8]{inputenc}
\usepackage[T1]{fontenc}
\usepackage{amsmath}

\shorttitle{3D reconnection in anisotropic turbulence}
\shortauthor{J.A. Agudelo Rueda and Others}

\title{Three-dimensional magnetic reconnection in particle-in-cell simulations of anisotropic plasma turbulence}

\author{Jeffersson A. Agudelo Rueda\aff{1}
  \corresp{\email{jeffersson.agudelo.18@ucl.ac.uk}},
  Daniel Verscharen\aff{1,2}, 
  Robert T. Wicks\aff{1,3},
  Christopher J. Owen\aff{1},
  Georgios Nicolaou\aff{1,7},
  Andrew P. Walsh \aff{5},
  Ioannis Zouganelis\aff{5},
  Kai Germaschewski\aff{2}, 
   \and Santiago Vargas Dom\'inguez \aff{6}}

\affiliation{\aff{1}Mullard Space Science Laboratory, University College London, Surrey, UK
\aff{2}Space Science Center, University of New Hampshire, Durham NH, USA
\aff{3}Department of Mathematics, Physics $\&$ Electrical Engineering, Northumbria University, Newcastle upon Tyne, NE1 8ST, UK
\aff{4}Southwest Research Institute, San Antonio, TX 78238, USA
\aff{5}European Space Astronomy Centre, European Space Agency, Spain
\aff{6}Observatorio Astronómico Nacional, Universidad Nacional de Colombia, Colombia
}
\begin{document}

\maketitle

\begin{abstract}
We use 3D fully kinetic particle-in-cell simulations to study the occurrence of magnetic reconnection in a simulation of decaying turbulence created by anisotropic counter-propagating low-frequency Alfvén waves consistent with critical-balance theory. We observe the formation of small-scale current-density structures such as current filaments and current sheets as well as the formation of magnetic flux ropes as part of the turbulent cascade. The large magnetic structures present in the simulation domain retain the initial anisotropy while the small-scale structures produced by the turbulent cascade are less anisotropic. To quantify the occurrence of reconnection in our simulation domain, we develop a new set of indicators based on intensity thresholds to identify reconnection events in which both ions and electrons are heated and accelerated in 3D particle-in-cell simulations. According to the application of these indicators, we identify the occurrence of reconnection events in the simulation domain and analyse one of these events in detail. The event is related to the reconnection of two flux ropes, and the associated ion and electron exhausts exhibit a complex three-dimensional structure. We study the profiles of plasma and magnetic-field fluctuations recorded along artificial-spacecraft trajectories passing near and through the reconnection region. Our results suggest the presence of particle heating and acceleration related to small-scale reconnection events within magnetic flux ropes produced by the anisotropic Alfvénic turbulent cascade in the solar wind. These events are related to current structures of order a few ion inertial lengths in size.
\end{abstract}

\section{Introduction}

The solar wind is a low-collisionality plasma produced in the solar corona \citep{marsch2006kinetic}. It expands across the solar system exhibiting spatial and temporal variations in composition, density, velocity and temperature as well as in the electric and magnetic fields. The solar wind shows a non-adiabatic temperature profile with distance from the Sun \citep{gazis1982voyager} which suggests the presence of local heating and particle-acceleration mechanisms \citep{goldstein2015kinetic}. Unlike in collisional plasmas, in the solar wind the energy dissipation cannot be attributed to the viscous interaction due to binary particle collisions nor to any process that depends directly on collisions, such as the collisional electric resistivity for instance. In the solar wind, the magnetic-field fluctuations exhibit a power-law distribution of the magnetic energy across a large range of spatial scales from 0.1 au to sub-proton scales \citep{coleman1968turbulence, marsch1990spectral} which indicates the presence of turbulence in the solar wind. The energy cascade has three regimes, the so-called injection range in which the power index of the magnetic-field fluctuations is $-1$ \citep{kiyani2015dissipation}, an inertial range in which the power index varies from $-3/2$ to $-5/3$ \citep{iroshnikov1963turbulence,marsch1990spectral,podesta2009dependence,boldyrev2011spectral} and a dissipation range in which the power index is less clearly defined \citep{goldstein1994properties,li2001dissipation,howes2008model} with spectral breaks at electron scales \citep{alexandrova2009universality, sahraoui2009evidence}. The transport of energy between scales is known as the energy cascade. At sub-proton scales, kinetic dissipation mechanisms become important, particles are energised and the entropy of the system irreversibly increases \citep{tatsuno2009nonlinear,eyink2018cascades, verscharen2019multi}. The nature of the fluctuations at sub-proton scales and the properties of the plasma determine whether the turbulent energy is mainly dissipated by ions or whether it cascades to electron scales at which it is ultimately dissipated by electrons. In the framework of wave turbulence, the energy-dissipation mechanisms are classified into two main categories: resonant heating such as Landau damping and ion-cyclotron damping \citep{marsch2003ion, kasper2008hot} and non-resonant heating such as stochastic heating \citep{chandran2010perpendicular,chandran2013stochastic}. In this framework, turbulent fluctuations with polarisation properties consistent with kinetic Alfvén waves (KAWs) and whistler waves are often evoked as the mechanisms that carry the turbulent cascade to electron scales. In general, observations more often find evidence for KAW-like fluctuations than for whistler-wave-like fluctuations \citep{smith2011observational,salem2012identification,podesta2012scale,podesta2013evidence,goldstein2015kinetic}. Another mechanism proposed to carry the turbulent cascade to sub-proton scales is magnetic reconnection \citep{sundkvist2007dissipation,franci2017magnetic, loureiro2020nonlinear}.
\\
\indent Magnetic reconnection is a process in which particles are heated and accelerated while the magnetic field topology changes. It takes place when magnetic structures form a region in which the frozen-in condition is locally broken allowing the exchange of particles between the magnetic structures and leading to a change in the magnetic connectivity \citep{hesse1988theoretical, pontin2011three}. Magnetic reconnection is a multiscale phenomenon that appears in both space and laboratory plasmas under conditions reaching from fully collisional to effectively collisionless. It has been predicted to occur in coronal mass ejections, solar flares, explosive events in planetary magnetospheres, accretion discs, star-formation regions, fusion plasmas and in the solar wind \citep[see][]{priest2007magnetic, zweibel2009magnetic}.
In the latter, reconnection events are characterised by streams of particles associated with Alfv\'enic disturbances and magnetic-field rotations \citep{gosling2005direct, davis2006detection,gosling2006petschek, phan2006magnetic, phan2009prevalence, gosling2012magnetic,phan2020parker}. These structures are interpreted as the so-called ``exhaust regions'' of reconnection events. Although magnetic reconnection has been studied for over 60 years, there is still no consensus in terms of a complete theory to describe magnetic reconnection at all scales involved. The problem is rooted in the fact that the range of spatial ($L$) and temporal ($\tau$) scales involves fluid-like behaviour at $L \gg  \rho_{i}, d_{i}$, where $\rho_i$ is the ion gyroradius and $d_{i}$ is the ion inertial length, as well as kinetic behaviour and energy dissipation at sub-proton scales, $L \ll  \rho_{i}, d_{i}$.  In addition, since plasmas are often in a turbulent state, the presence of a turbulent field alters the onset and evolution of reconnection events \citep{matthaeus1986turbulent, strauss1988turbulent, lazarian1999reconnection, kim2001turbulent, servidio2011magnetic, boldyrev2017magnetohydrodynamic, adhikari2020reconnection, loureiro2020nonlinear}. It is unclear how turbulence and reconnection affect each other and how the energy is partitioned between particles and fields through both processes. For instance, although the role of reconnection in the small-scale turbulent cascade has been studied previously \citep{franci2017magnetic, boldyrev2017magnetohydrodynamic, cerri2017reconnection, papini2019fast}, it is still unclear how 3D reconnection proceeds in the turbulent solar wind. It is not well understood whether 3D reconnection disrupts current sheets and coherent magnetic-field structures associated with intermittency at small scales in the same way as it disrupts these structures at large scales \citep{boldyrev2013toward, mallet2017disruption}. Moreover, it is unclear how reconnection changes the turbulent cascade as the wavevector anisotropy increases with decreasing scale and how turbulence affects the reconnection process itself \citep{boldyrev2017magnetohydrodynamic}. Therefore, it is necessary to study the energy partition as well as the links between turbulence and reconnection at small scales in order to fully understand the mechanisms of energy dissipation and plasma heating in the solar wind. 
\\
\indent The use of numerical simulations has been proven to be an invaluable tool to test existing theories over a wide range of parameters. Moreover, using simulations, we self-consistently explore nonlinear problems which lie beyond analytical theory. Simulations expand our knowledge regarding magnetic reconnection processes in 2D \citep{birn2001geospace, shay2001alfvenic, servidio2009magnetic, loureiro2009turbulent, Servidio2010, bessho2017effect} and in 3D \citep{hesse2001particle, pritchett2001kinetic, lapenta2003new, lapenta2006kinetic, kowal2009numerical, daughton2011role, pritchett2013onset, baumann20133d, liu2013bifurcated, munoz2018kinetic}. The use of high-performance computing facilities and the increasing computational capabilities facilitate the study of plasmas from first principles using particle-in-cell (PIC) simulations \citep{lapenta2012particle, germaschewski2016plasma}. These simulations are able to resolve proton and electron scales and to account for phenomena that only reveal themselves using kinetic theory. For instance, electron-kinetic effects can affect ion-scale processes \citep{told2016comparative} even in linear theory. These effects may be even enhanced in nonlinear processes. Currently, full PIC simulations are unable to cover the whole range of scales involved in natural plasma turbulence and reconnection since they are expensive in terms of computing memory and require small time steps to satisfy stability criteria. However, their ability to model the physics behind the energy partition at small scales makes PIC the most appropriate method to address sub-proton and electron-scale phenomena as well as collisionless energy dissipation. 
\\
\indent Kinetic simulations of magnetic reconnection are often based on idealised conditions, such as the Harris current-sheet configuration \citep{shay2001alfvenic, scholer2003onset, shay2004scaling, Ricci2004, daughton2006fully, daughton2011role, liu2013bifurcated,  Leonardis2013, goldman2016can, beresnyak2016three}. In this work, we study the formation of current structures and the occurrence of 3D magnetic reconnection as a result of turbulent dynamics in PIC simulations of collisionless anisotropic Alfvénic turbulence. We initialise our simulation with counter-propagating Alfv\'en waves that then self-consistently interact and generate turbulence \citep{howes2013alfven1, howes2015dynamical}, current-sheet structures \citep{howes2016dynamical} and regions of magnetic reconnection. The overall objective of this work is to discover the properties of reconnection events that terminate the inertial-range cascade of solar-wind turbulence and define criteria that can identify such features in future 3D simulations and in spacecraft data. These results will allow future work to advance the study of linked reconnection and turbulence based on a solid and consistent framework of observable features. In Section \ref{sec:Initialization_sim}, we describe our initial conditions for the simulation as well as our numerical setup. We present our results in Section \ref{sec:discusison} and our conclusions in Section \ref{sec:conclusions}. 

\section{Simulation}
\label{sec:Initialization_sim}

We use the explicit Plasma Simulation Code \citep[PSC,][]{germaschewski2016plasma} to simulate eight anisotropic counter-propagating Alfvén waves in an ion-electron plasma. Since the theories of turbulence dissipation through reconnection in the solar wind are intrinsically connected to anisotropy through the generation of thin structures that form the precursors of current sheets, our initial waves are anisotropic. The anisotropy of the initial fluctuations is set up according to the theory of critical balance by \citet{sridhar1994toward} and \citet{goldreich1994toward}, henceforth GS95. A detailed explanation of the initial conditions is presented in Appendix \ref{app:initial}. The normalization parameters are the speed of light $c$, the vacuum permittivity $\epsilon_{0}$, the magnetic permeability $\mu_{0}$, the Boltzmann constant $k_{b}$, the elementary charge $q$, the ion mass $m_{i}$, the initial density of ions and electrons $n_{i}=n_{e}$ and the ion inertial length $d_{i}=c/\omega_{pi}$ where $\omega_{pi}=\sqrt{n_{i}q^{2}/m_{i}\epsilon_{0}}$ is the ion plasma frequency. We set $\beta_{s,\parallel}=1$ and $T_{s,\parallel}/T_{s,\perp}=1$, where $\beta_{s,\parallel}=2 n_s \mu_{0} k_{B}T_{s,\parallel}/B_{0}^{2}$ is the parallel beta, $\mathbf{B}_{0}$ is the background magnetic field, and the index $s$ indicates the plasma species. $T_{s,\parallel}$ and $T_{s,\perp}$ are the parallel and perpendicular temperatures respectively. The magnetic field is normalised to the value of the constant background field $B_0$ and the Alfvén speed ratio is $V_{A}/c=0.1$, where $V_{A}=B_{0} / \sqrt{\mu_{0}n_{i}m_{i}}$ is the ion Alfvén speed. We use $100$ particles per cell ($100$ ions and $100$ electrons), a mass ratio of $m_{i}/m_{e} = 100$ so that $d_e = 0.1 d_{i}$ where $m_{e}$ is the electron mass and $d_{e}$ is the electron inertial length. The simulation box size is $L_{x} \times L_{y} \times L_{z} = 24d_{i}\times24d_{i}\times125d_{i}$ and the spatial resolution is $\Delta x =\Delta y = \Delta z =  0.06d_{i}$. We use a time step of $\Delta t =0.06/ \omega_{pi}$. In our normalisation, the Debye length $\lambda_{D}=d_{i}\sqrt{\beta_{i}/2}V_{A}/c$ defines the minimum spatial distance that needs to be resolved in the simulation and $\lambda_D=0.07d_i$.  Although our numerical parameters $V_A/c$ and $m_i/m_e$ are not identical to the corresponding parameters in the solar wind, they allow us to perform simulations within the computational limitations. With these parameters, the simulated electrons are mildly relativistic, which they are not in the real solar wind. However, the effect of mildly relativistic electrons on the propagation and damping of kinetic-scale normal modes, including kinetic Alfvén waves (KAWs), Alfvén/ion-cyclotron (A/IC) waves and fast-magnetosonic/whistler (FM/W) waves, is negligible \citep{verscharen2020dependence} and not important for the evolution of the turbulent cascade regardless of the processes that carry the cascade to subproton scales. 


\section{Results}
\label{sec:discusison}

In this section, we discuss the time evolution (Section \ref{subsec:time_evolution}) and the spectral properties of the turbulence in our simulation (Section \ref{subsec:evidence_turbulence}). We then define a new set of indicators of reconnection based on 2D and 3D reconnection models and study a self-consistently formed reconnection region in detail (Section \ref{subsec:finding_sites}). We then record and discuss the plasma properties that an artificial spacecraft observes in the spacecraft frame as it passes through our simulation box (Section \ref{subsec:1d_trajectories}). 

\subsection{Time evolution and formation of current structures}
\label{subsec:time_evolution}

\begin{figure}
\centering
\includegraphics[width=0.7\textwidth]{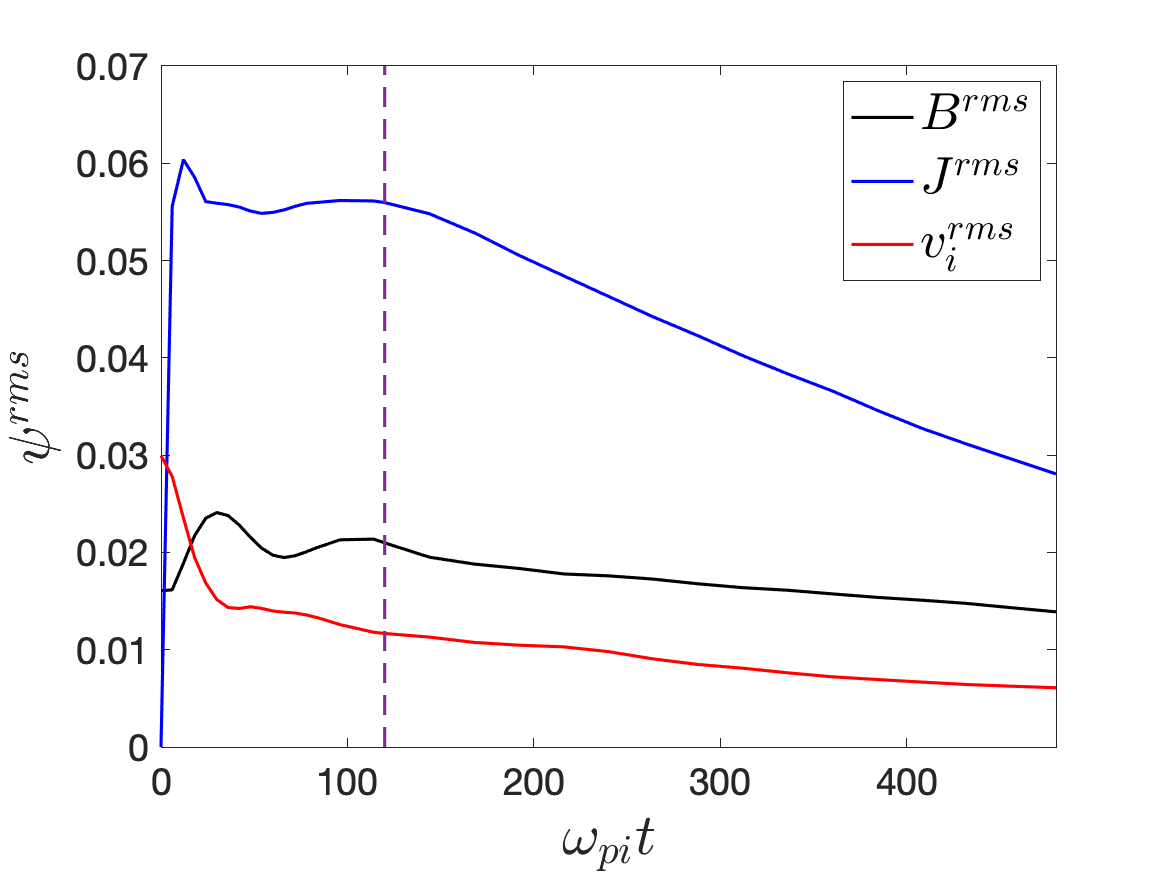}
\caption{Time evolution of the rms of the current density $\mathbf{J}$ (blue), magnetic field $\mathbf{B}$ (black) and ion velocity $\mathbf{v}_{i}$ (red). The vertical dashed line marks the time $t_{R} = 120/\omega_{pi}$ at which $J^{rms}$ begins to decrease.  }
\label{fig:JBVIrms}
\end{figure}

We first identify a representative time $t_R$ for our subsequent analysis of the turbulence properties. The root mean square (rms) of the current density $J^{rms}$ is an indicator commonly used to identify the time at which the system reaches a quasi-stationary state. At this time, the generation of current sheets by waves is balanced by their decay so that the growth of $J^{rms}$ saturates, which marks the time of maximum turbulent activity in the simulation \citep{franci2017magnetic}. The rms of a quantity $\psi$ is defined as

\begin{align}
    \psi^{rms} = \sqrt{\langle \psi^{2} \rangle - \langle \psi \rangle^{2}},
\end{align}

\noindent where $\langle ... \rangle$ represents the spatial average over the whole simulation domain. Figure \ref{fig:JBVIrms} shows the time evolution of the rms of the current density $\mathbf{J}$ (blue), the magnetic field $\mathbf{B}$ (black) and the ion velocity $\mathbf{v}_{i}$ (red) in our simulation. Since we start our simulation under the assumption that the linear time $\tau_{l}$ is approximately equal to the nonlinear time $\tau_{nl}$, we estimate  $\tau_{nl} \sim \tau_{l} \sim 1/k_{\parallel}V_{A} \sim L_{z}/2\pi V_{A} \approx 200/\omega_{pi}$. This estimate for the nonlinear time $\tau_{nl}$ is therefore related to the scale of the initial fluctuations and represents an upper limit. We observe a peak in $J^{rms}$ at $t=12/\omega_{pi}$ which is due to the self-consistent formation of current structures as a response to the initial magnetic-field fluctuations. The variation in  $B^{rms}$ and $J^{rms}$ during the initial phase, between $t=12/\omega_{pi}$ and $t=96/\omega_{pi}$, suggests that the system is still in a phase of self-adjustment. The formation of the plateau in $J^{rms}$  at $t \approx \tau_{nl}/2 \approx 100/\omega_{pi}$ indicates that the system has reached a quasi-stationary state. Therefore, we expect the formation of current structures such as current sheets and current filaments by this time. The vertical dashed line marks the time $t = 120/\omega_{pi}$ at which $J^{rms}$ begins to decrease monotonously until the simulation ends. In this sense, the time $t=120/\omega_{pi}$ represents the beginning of the decaying phase in our system. As the system evolves in time, current and magnetic structures dissipate, and we expect an exchange of the energy stored in the magnetic field with the kinetic energy of the particles. Based on these considerations, we use the time $t_{R}=120/\omega_{pi}$ to study the spectral properties of the turbulence in our system.  

\begin{figure}
\centering
\begin{subfigure}[b]{0.9\linewidth}
\includegraphics[width=\textwidth]{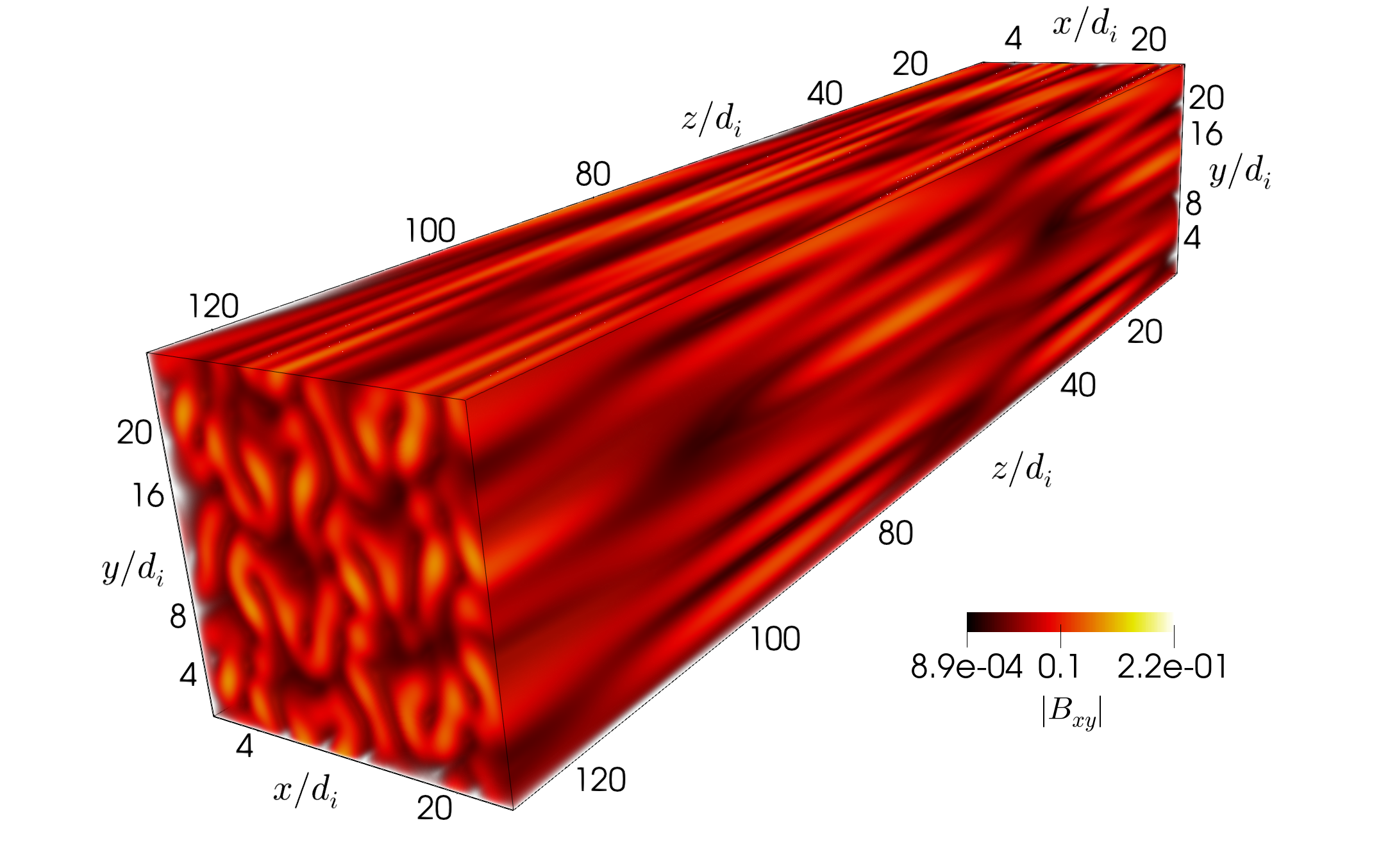}
\caption{}
\label{fig:simut12a}
\end{subfigure}
\begin{subfigure}[b]{0.9\linewidth}
\includegraphics[width=\textwidth]{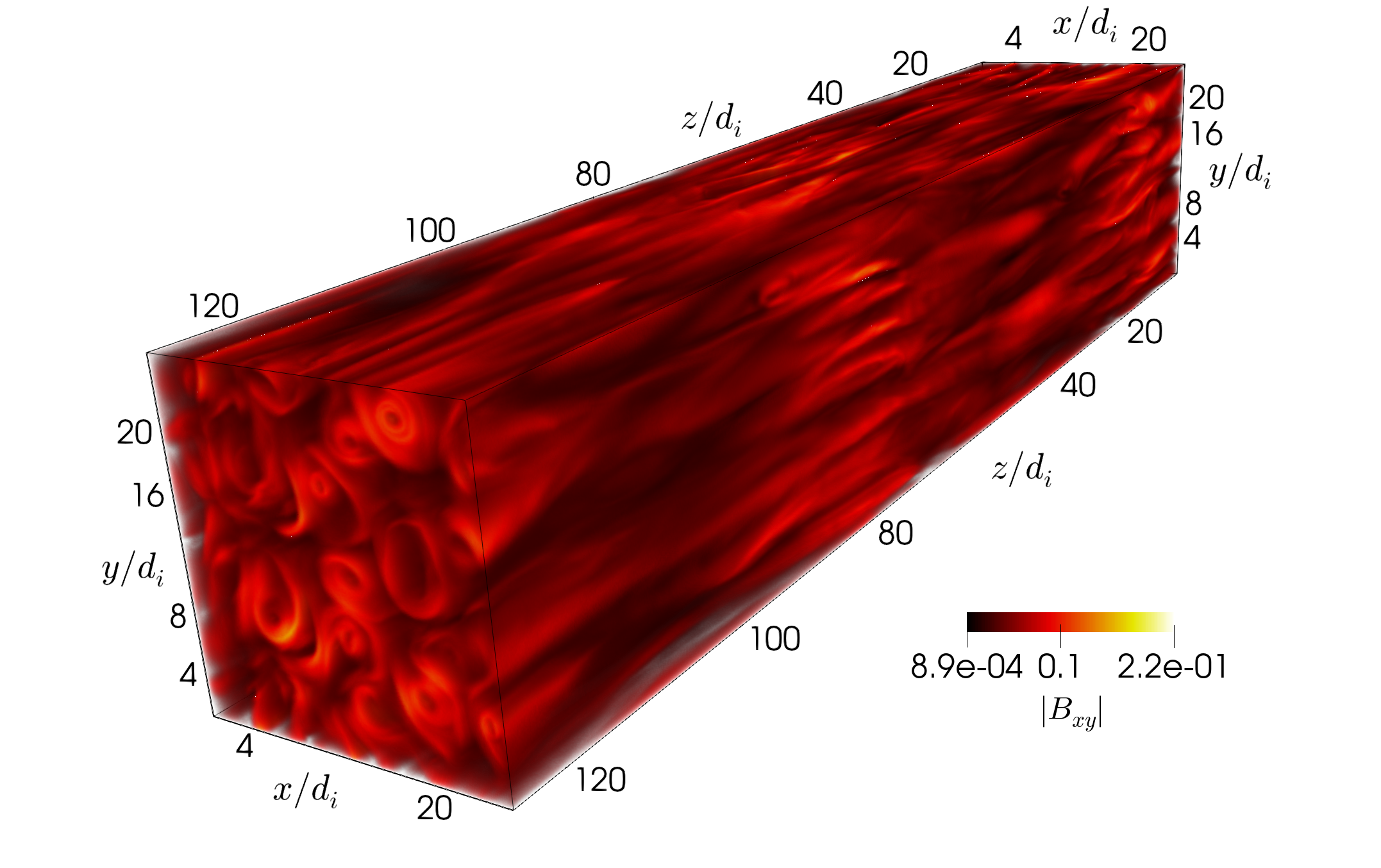}
\caption{}
\label{fig:simut12b}
\end{subfigure}
\caption{3D rendering of the transverse magnetic field magnitude $|\mathbf{B}_{xy}| = \sqrt{B_{x}^{2}+B_{y}^{2}}$ at $t=0$ (\textit{a}) and $t=t_{R}$ (\textit{b}). The colour bar ranges from the minimum magnitude (black) to the maximum magnitude (yellow) throughout the simulation domain at $t=t_{R}$. We use the same colour bar in both panels for a direct comparison. The initial background magnetic field is directed along $z$-direction. At the initial time, the fluctuations are anisotropic and elongated along the $z$-direction. At $t=t_{R}$, small-scale magnetic eddies have formed and interact nonlinearly with each other. The eddies present varying cross section diameters $L_{D}$ and lengths $L_{\parallel}$.}
\label{fig:simut12}
\end{figure}

\begin{figure}
\centering
\begin{subfigure}[b]{0.47\linewidth}
\includegraphics[width=\textwidth]{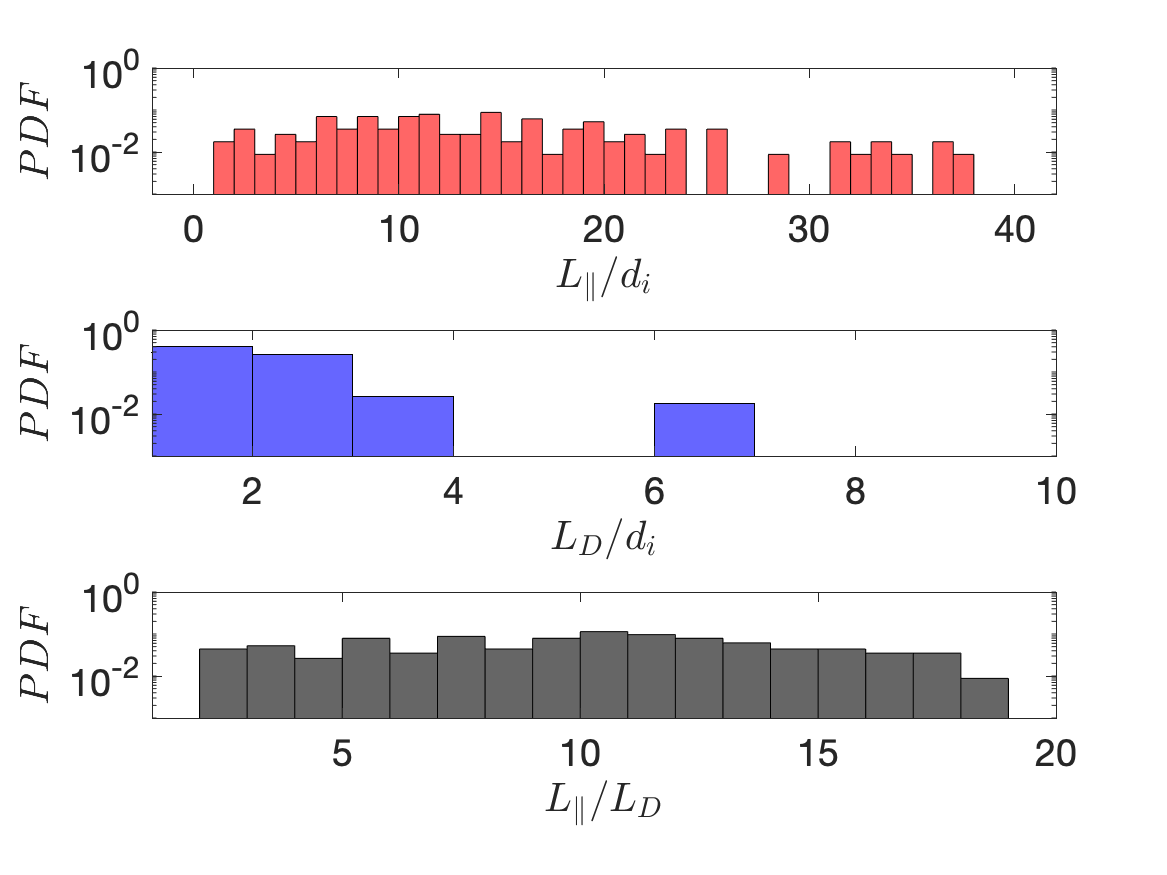}
\caption{$t=0$}
\label{fig:shapes0}
\end{subfigure}
\begin{subfigure}[b]{0.47\linewidth}
\includegraphics[width=\textwidth]{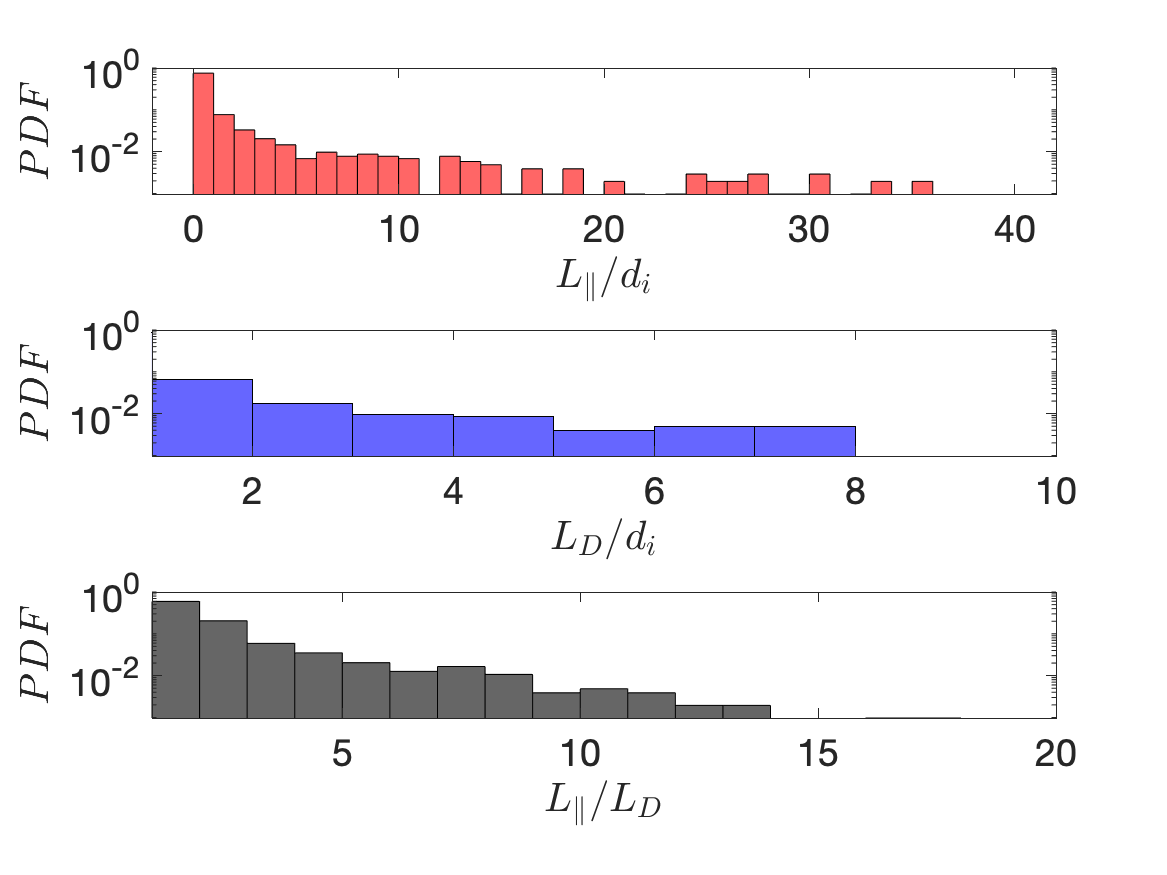}
\caption{$t=t_{R}$}
\label{fig:shapes120b}
\end{subfigure}
\begin{subfigure}[b]{0.47\linewidth}
\includegraphics[width=\textwidth]{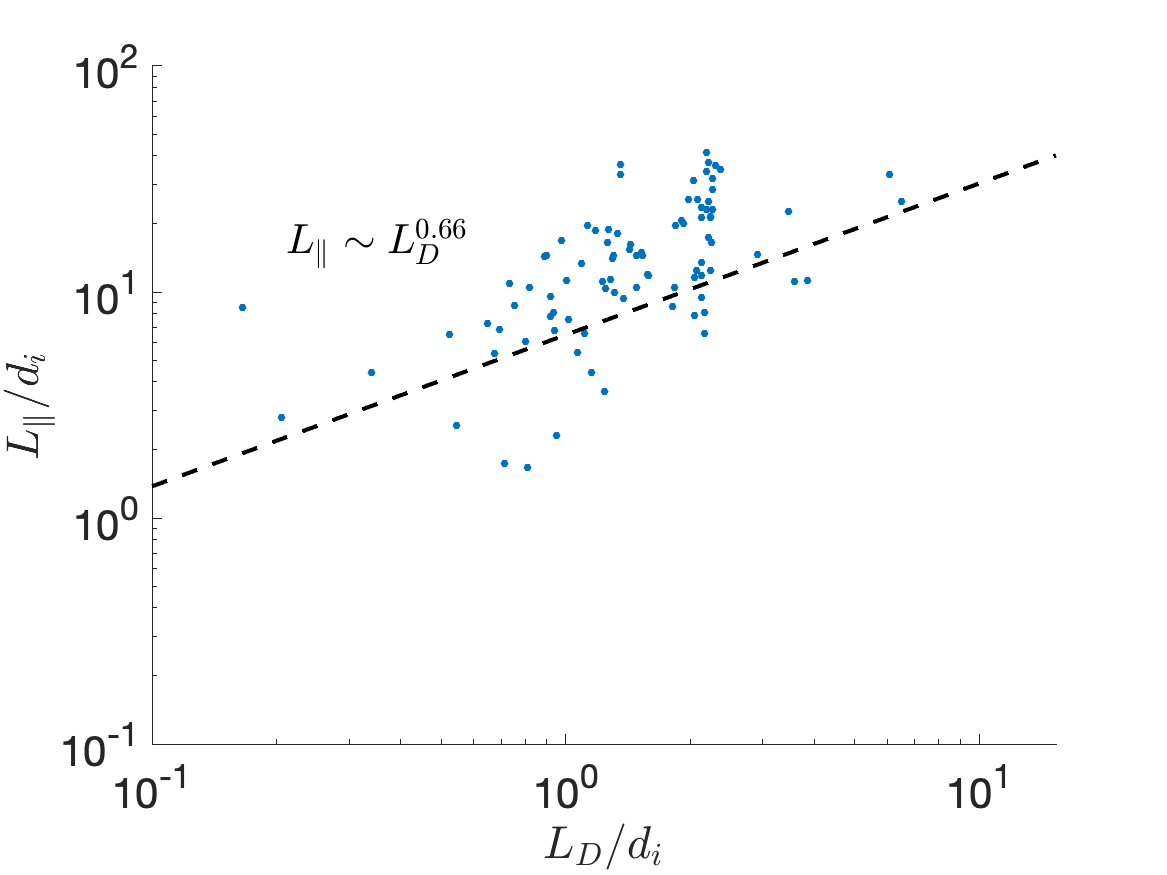}
\caption{$t=0$}
\label{fig:shapes1}
\end{subfigure}
\begin{subfigure}[b]{0.47\linewidth}
\includegraphics[width=\textwidth]{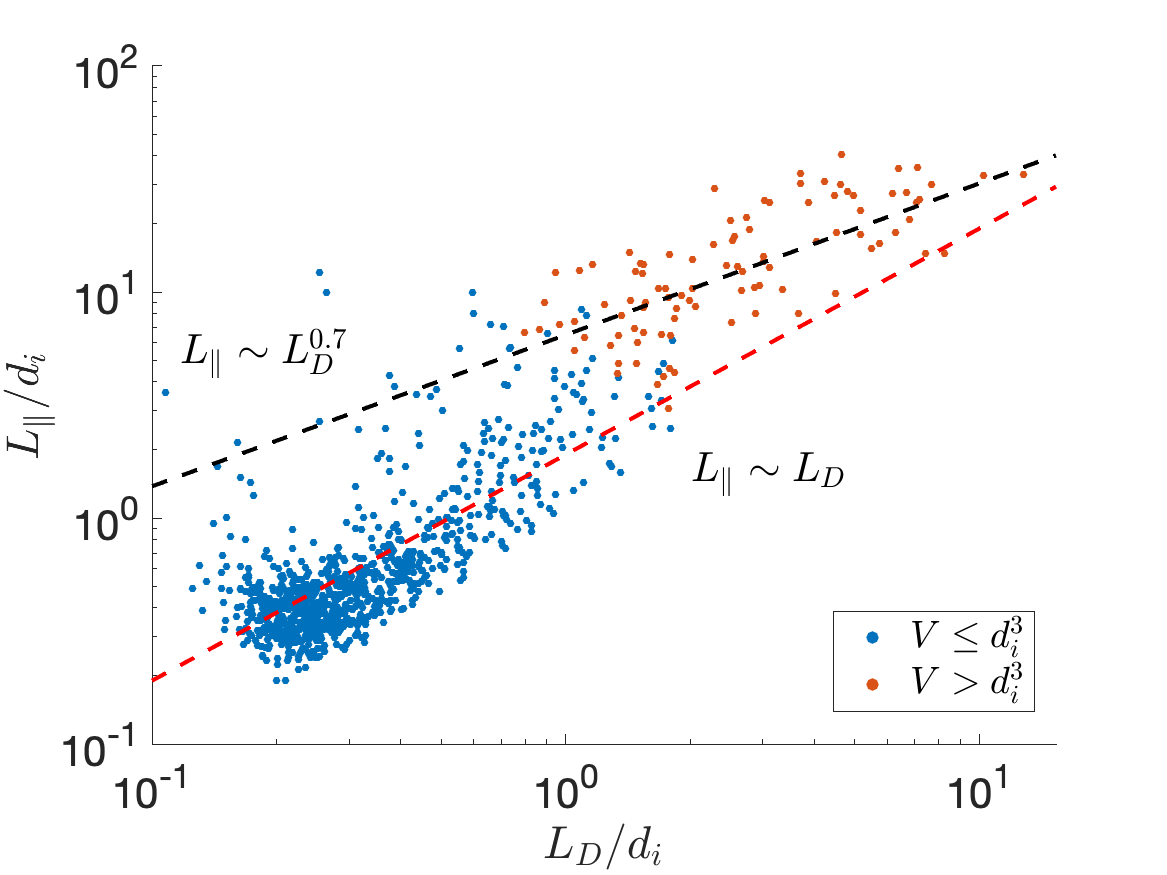}
\caption{$t=t_{R}$}
\label{fig:shapes120d}
\end{subfigure}
\caption{Panels (a) and (b): Probability distribution functions of elongations $L_{\parallel}$ (top), cross section diameters $L_D$ (middle) and aspect ratios $L_{\parallel}/L_D$ (bottom) of the magnetic structures at $t=0$ (a) and $t=t_R$ (b). Panel (c): Scaling between $L_{\parallel}$ and $L_D$ at $t=0$. The black dashed line show the linear fit. Panel (d): Scaling between $L_{\parallel}$ and $L_D$ of the large-scale population (orange) and small-scale population (blue) at $t=t_{R}$. The top black dashed line shows the linear fit to the former while the bottom red dashed line shows linear fit to the latter.}
\label{fig:shapes}
\end{figure}

\begin{figure}
\centering
\begin{subfigure}[b]{0.9\linewidth}
\includegraphics[width=\textwidth]{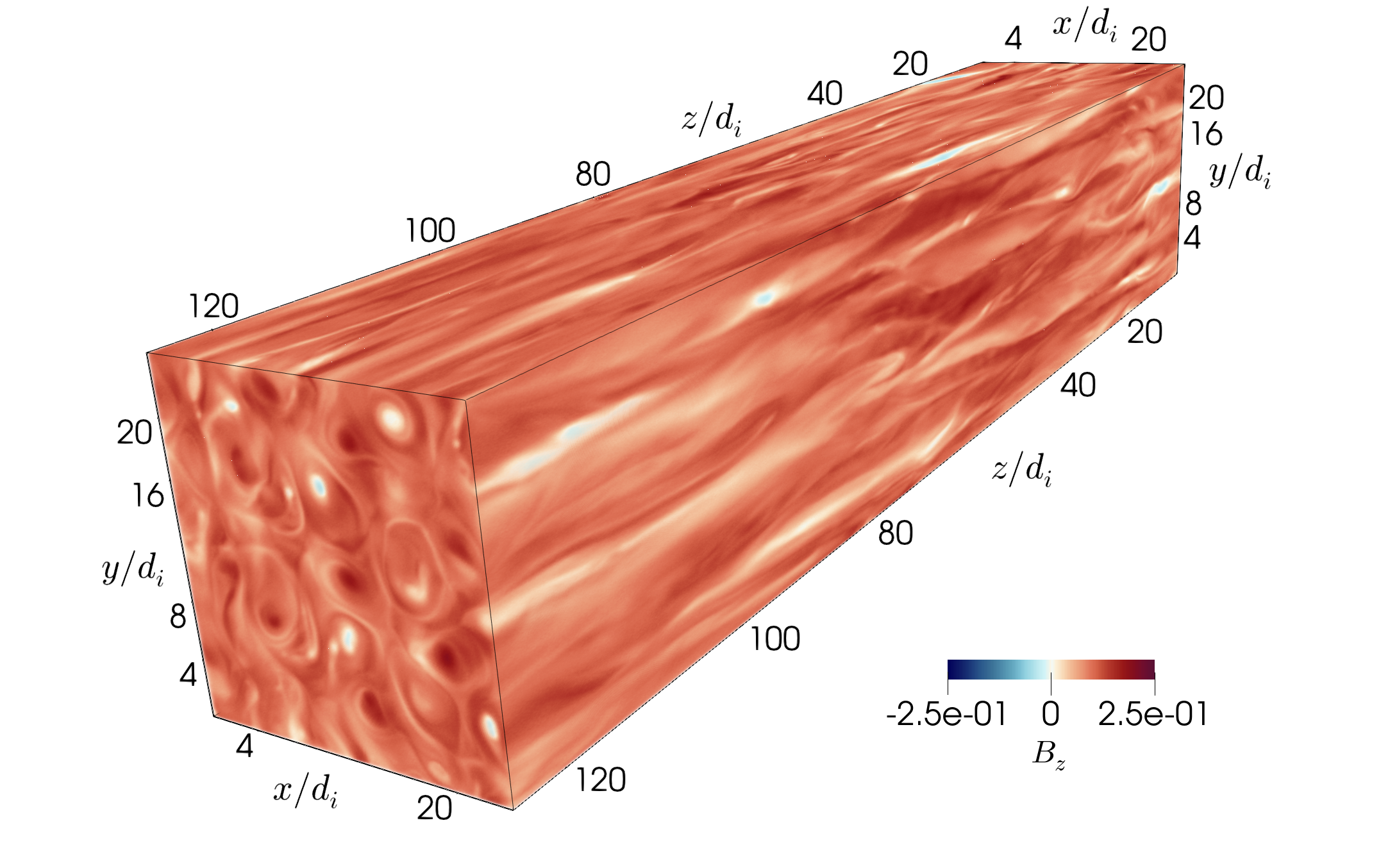}
\caption{}
\label{fig:simut34a}
\end{subfigure}
\begin{subfigure}[b]{0.9\linewidth}
\includegraphics[width=\textwidth]{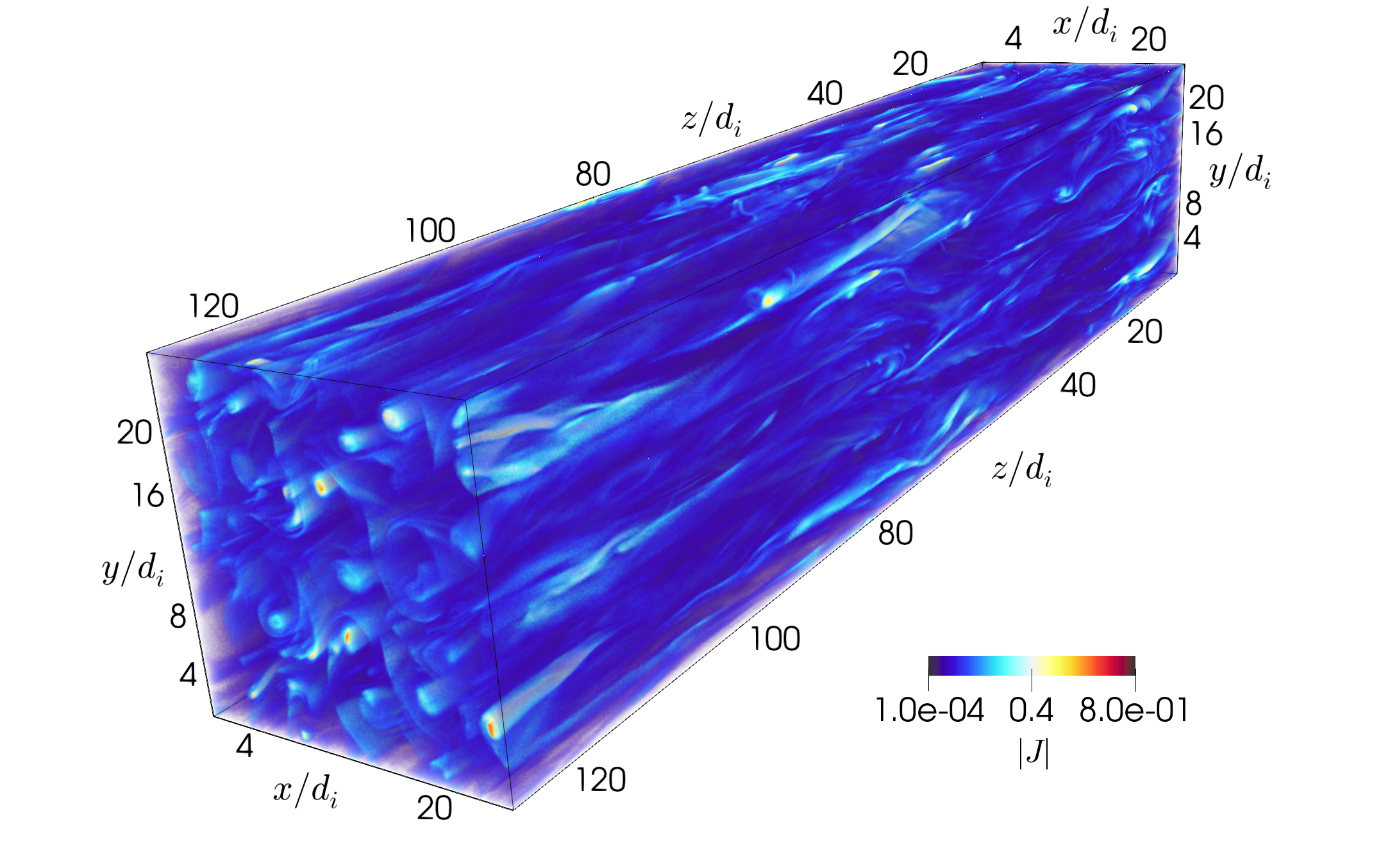}
\caption{}
\label{fig:simut34b}
\end{subfigure}
\caption{Visualisations of the simulation domain at $t=t_{R}$. (\textit{a}) 3D rendering of the magnetic-field component $B_{z}$.  Blue represents negative, red positive and white zero values of $B_{z}$. The eddies' centres present different values of $B_z$ with either positive or negative polarity. (\textit{b}) 3D rendering of the magnitude of the current density $| \mathbf{J} |$ from the same vantage point as (\textit{a}). The colour represents in blue (red) the smallest (largest) values of $|\mathbf{J}|$. Filaments of intense current density are aligned with the eddies' centres. Current filaments and extended current-sheet-like structures are mainly elongated along the $z$-direction.}
\label{fig:simut34}
\end{figure}

Figure \ref{fig:simut12} shows a 3D rendering of the magnitude of the transverse magnetic field $|\mathbf{B}_{xy}| = \sqrt{B_{x}^{2}+B_{y}^{2}}$ at two different time steps: panel (\textit{a}) at $t=0$ and panel (\textit{b}) at $t=t_{R}$. Panel (\textit{a}) shows the anisotropic interference pattern of the linear superposition of Alfv\'en waves at $t=0$. Initially, there are no coherent eddies present because no nonlinear interaction has taken place yet. However, the initial magnetic-field fluctuations are already anisotropic. Panel (\textit{b}) shows that at time $t=t_{R}$, there is a clear presence of magnetic eddies with varying cross section diameters $L_{D}$ and elongations $L_{\parallel}$, where $L_{\parallel}$ represents the length of these eddies along the local magnetic field. Even though we start with a superposition of only eight waves, nonlinear interactions generate magnetic eddies of different shapes and anisotropies. At this time, the magnetic-field structures consist of a combination of linear fluctuations and magnetic eddies. To estimate the shape of the magnetic structures at $t=0$ and $t=t_{R}$, we calculate $\Delta B=\sqrt{B_{x}^{2}+B_{y}^{2}+(B_{z}-B_{0})^{2}}$ and use an intensity threshold defined as $\Delta B > \langle \Delta B \rangle+2 \Delta B^{rms}$. We define a magnetic structure as the combination of those cells in our simulation that are connected as next neighbours and fulfil this threshold condition. The exact value of the threshold is chosen to improve the performance of the algorithm in the identification of these structures. After the identification of the structures, we calculate their principal axes. We define $L_D=\sqrt{L_{\perp 1}^2+L_{\perp 2}^2}$, where $L_{\perp 1}$ and $L_{\perp 2}$ are the two orthogonal diameters in the plane perpendicular to the local magnetic field and $L_{\parallel}$ is the axis along the local magnetic field. 

\indent Panel (\textit{a}) of Figure \ref{fig:shapes} shows the probability distribution function ($PDF$) of $L_{\parallel}$, $L_{D}$ and the aspect ratio $L_{\parallel}/L_{D}$ at $t=0$ and panel (\textit{b}) at $t=t_{R}$. The mean value and standard deviation of the distributions of $L_{\parallel}$, $L_D$, and $L_{\parallel}/L_D$ at $t=0$ are ${L}_{\parallel} = (16.33 \pm 8.32)d_{i}$, ${L}_{D} = (1.55 \pm 0.95)d_{i}$ and ${\left(L_{\parallel}/L_{D}\right)} = (11.01 \pm 7.06)d_{i}$. At $t=t_{R}$, we find ${L}_{\parallel} = (2.16 \pm 5.08)d_{i}$, ${L}_{D} = (0.62 \pm 0.72)d_{i}$ and ${L_{\parallel}/L_{D}} = (2.55 \pm 1.94)d_{i}$. This shows that the nonlinear interaction has formed magnetic structures with smaller elongations and cross section diameters continuously distributed between $L_{D}=1d_{i}$ and $8 d_{i}$. The distribution of aspect ratios is less uniform at $t=t_{R}$ than at $t=0$. The number of magnetic structures with nearly isotropic aspect ratios is greater at $t=t_{R}$. To study the distribution of the large-scale structures at $t=t_{R}$, we further apply a filter to remove all regions with an equivalent volume $V \leq 1d_{i}^{3}$, where $V$ is defined as the space filled by the sum of all contiguous cells associated with a given magnetic structure. For all structures with $V>d_{i}^{3}$, we find ${L}_{\parallel} = (14.97 \pm 9.01)d_{i}$, ${L}_{D} = (3.14 \pm 2.25)d_{i}$ and ${L_{\parallel}/L_{D}} = (5.46 \pm 2.48)d_{i}$. The distribution of the large-scale magnetic structures maintains an anisotropy consistent with our initial conditions. Panel (\textit{c}) of Figure \ref{fig:shapes} shows the scaling between $L_{\parallel}$ and $L_{D}$ for the magnetic structures at $t=0$. The linear fit to these structures, dashed line, reveals the scaling $L_{\parallel} \sim L_{D}^{0.66}$ which is consistent with our initial anisotropy, i.e., $L_{\parallel} \sim L_{D}^{2/3})$. Panel (\textit{d}) of Figure \ref{fig:shapes} shows the scaling between $L_{\parallel}$ and $L_{D}$ for the magnetic structures at $t=t_{R}$. The orange dots represent the structures satisfying $V > d_{i}^{3}$ while the blue dots show the structures satisfying $V \leq d_{i}^{3}$. The linear fit to the former population, top black dashed line, reveals the scaling $L_{\parallel} \sim L_{D}^{0.7}$. In contrast, the linear fit to the latter population, bottom red dashed line, shows the isotropic scaling, $L_{\parallel} \sim L_{D}$. Around $L_{D} \sim d_{i}$ we find a transition and mixing between structures with both scalings. This suggests that the large-scale structures tend to maintain the initial anisotropy while the small-scale structures become more isotropic. This isotropic scaling at sub-proton scales has also been observed in hybrid simulations \citep{franci2018solar, arzamasskiy2019hybrid, landi2019spectral}.
\\
\indent Figure \ref{fig:simut34} shows 3D renderings of $B_{z}$ and $|\mathbf{J}|$ at $t=t_{R}$. Panel (\textit{a}) shows $B_{z}$, from the same vantage point as panel (\textit{b}) of Figure \ref{fig:simut12}. Although the initial $\mathbf{B}_{0}$ is uniform and points into the $+z$-direction, nonlinear interactions generate regions in which $B_{z}$ is negative. These regions are mostly localised in the centres of the small eddies in panel (\textit{b}) of Figure \ref{fig:simut12}. Panel (\textit{b}) in Figure \ref{fig:simut34} shows that the locations of the most intense current filaments coincide with the centres of the magnetic eddies. Current filaments are intense quasi-cylindrical current structures. Similar to the case of the magnetic structures, we apply the threshold $|J| \geq \langle |J|\rangle + 4(|J|)^{rms}$ to determine the shape of the current filaments. The mean cross section diameter of these current filaments is $\hat{L}_{D} = (1.94 \pm 0.84)d_{i}$. Their mean length is $\hat{L}_{\parallel} = (12.32 \pm 6.70)d_{i}$ and their mean aspect ratio is $\hat{L}_{\parallel}/\hat{L}_{D} = (6.84 \pm 3.48)$. The filaments are mostly elongated along the ${z}$-direction. Some filaments have undergone bending and twisting due to the nonlinear interactions. The elongations of the current filaments are distributed similarly to the elongations of the magnetic eddies (not shown here) and vary in the range of scales from  $\sim 4 d_{i}$ to $\sim 30 d_{i}$. Panel (\textit{b}) shows in addition the formation of thin current-sheet-like structures at the edges of the eddies where the perpendicular component of the magnetic field is nearly zero (see panel (\textit{b}) in Figure \ref{fig:simut12}). We define current sheets as current structures in which $L_{cs} \gg \delta_{cs}$ and $\Delta_{cs} \gg \delta_{cs}$, where $L_{cs}$ is the current-sheet length along the local magnetic field, $\Delta_{cs}$ is the current-sheet width tangential to the magnetic eddies and $\delta_{cs}$ is the current-sheet thickness normal to the edge of the eddies. The formation of these current sheets is due to the turbulent motions that squeeze the eddies together. In the supplementary material we provide a movie that shows the time evolution of the volume rendering of $J_{z}$ in the $zx$-plane. We observe the tearing and breaking up of current sheets as well as the onset of instabilities arising from the nonlinear interactions and of jets oblique to the major axes of the current sheets as a result of the turbulent evolution. However, a detailed study of these phenomena is beyond the scope of this manuscript.  



\subsection{Evidence of turbulence}
\label{subsec:evidence_turbulence}

A broad power-law spectrum of the fluctuations indicates the presence of turbulence as the energy cascades from large to small scales. To analyse the spectral properties of the system, following \cite{franci2018solar}, we calculate the energy associated with the 3D Fourier modes $\psi_{3D}(\mathbf{k})$ of a quantity $\psi$ as  

\begin{align}
    \psi_{3D}(\mathbf{k}) =   \Tilde{\psi}(\mathbf{k})\Tilde{\psi}^{*}(\mathbf{k}),
\label{eqn:3Dfour}
\end{align}

\noindent where $\mathbf{k}$ is the wavevector, $\Tilde{\psi}(\mathbf{k})$ is the 3D spatial Fourier transform of $\psi$ and $\Tilde{\psi}^{*}(\mathbf{k})$ represents its complex conjugate. If $\psi$ is a vector quantity, the 3D Fourier transform is taken over each component and the product is defined as

\begin{align}
\Tilde{\psi}(\mathbf{k})\Tilde{\psi}^{*}(\mathbf{k}) = \sum_{i}\Tilde{\psi}_{i}(\mathbf{k})\Tilde{\psi}_{i}^{*}(\mathbf{k}),
\end{align}

\noindent where the index $i$ represents the components $x,y$ and $z$. Since our system does not include any anisotropy within the plane perpendicular to the background magnetic field on average, we assume that the energy distribution in the turbulent fluctuations remains axially symmetric on average. Thus, the wavevector can be expressed, without loss of generality, as its perpendicular and parallel components $(k_{\perp}, k_{\parallel})$. We note that the local (rather than the global) average magnetic field defines the cylindrical symmetry axis for the turbulent fluctuations \citep{cho2000anisotropy}. However, we use the global background magnetic field as a proxy. This simplification is motivated by the strong alignment of the eddies with the background magnetic field at this time in our simulation (see Figures \ref{fig:simut12} and \ref{fig:simut34}). Moreover, the definition of the local magnetic field is a matter of ongoing research and debate \citep{podesta2009dependence,  chen2011anisotropy, tenbarge2012interpreting, oughton2015anisotropy, gerick2017uncertainty} and the development of an anisotropic energy cascade is sufficient for the determination of reconnection events in the present study.\footnote{An analysis of the fluctuations with respect to the local magnetic field based on second-order structure functions supports this assumption and is provided in Appendix \ref{app:sosf}.} Thus, we calculate the perpendicular and parallel components of the wavevector as $k_{\perp}=\sqrt{k_{x}^{2}+k_{y}^{2}}$ and $k_{\parallel}=k_{z}$, respectively, and assume that the fluctuations are statistically independent of the azimuthal angle. We integrate $\psi_{3D}$ over concentric rings in $k_\perp$-space. The energy associated with the $j$th-ring is 

\begin{align}
\psi^{j}_{2D}(k_{\perp},k_{\parallel}) =  \int_{k_{\perp}^{j}}^{k_{\perp}^{j}+ dk_{\perp}} \psi_{3D}(k_{\perp}, k_{\parallel}) 2\pi k_{\perp}^{j'}dk_{\perp}^{'},
\label{eqn:2Dfour_1} 
\end{align}

\noindent where the thickness $dk_{\perp}$ of these rings is taken as the magnitude of the smallest perpendicular wavevector in our system $dk_{\perp} = 2\pi / \sqrt{2}L_{x}$. To visualise the energy cascade in $k$-space as well as the level of anisotropy in the system, we compute the reduced 2D power spectral density $P^{\psi}_{2D}(k_{\perp},k_{\parallel})$ as 

\begin{align}
 P^{\psi}_{2D}(k_{\perp},k_{\parallel}) = \sum_{j}  \frac{1}{k_{\perp}}\psi^{j}_{2D}(k_{\perp},k_{\parallel}).
\end{align}

\begin{figure}
\centering
\begin{subfigure}[b]{0.49\linewidth}
\includegraphics[width=\textwidth]{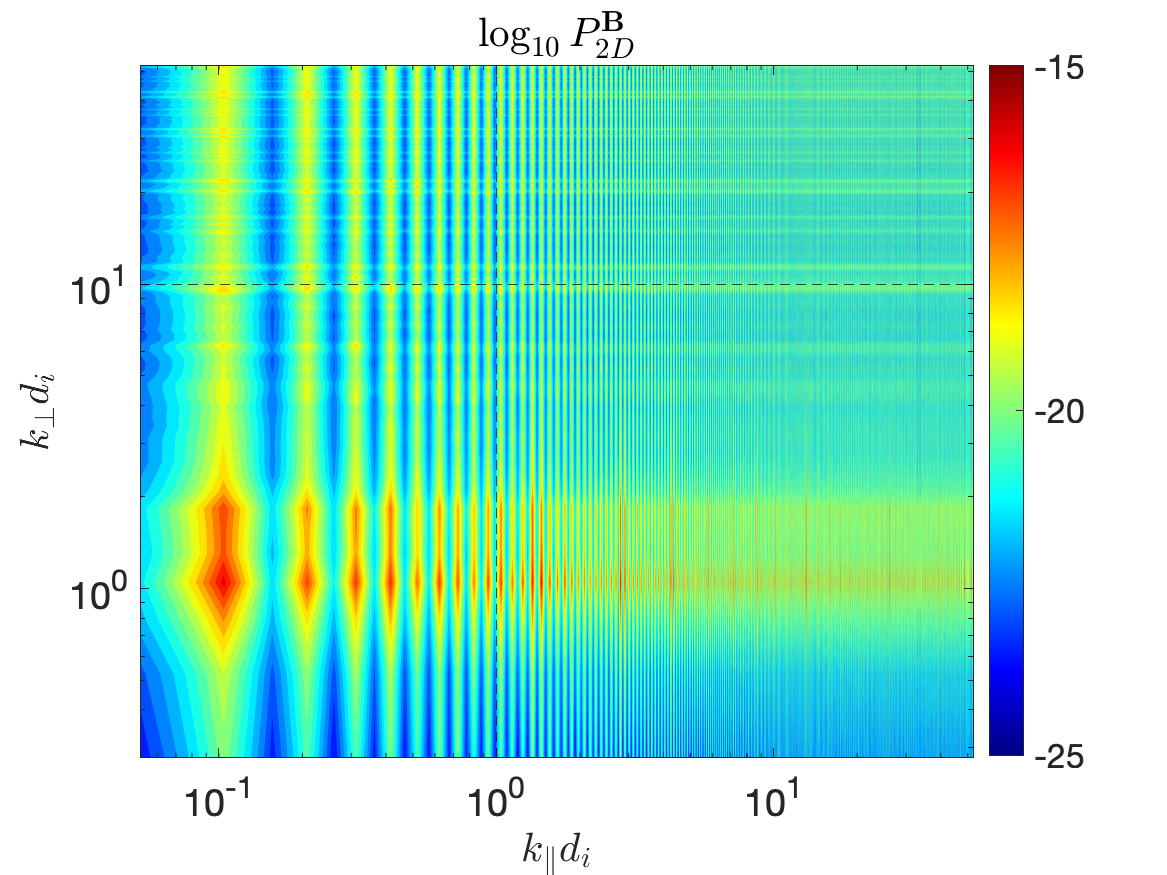}
\caption{$t=0/\omega_{pi}$}
\label{fig:2D12a}
\end{subfigure}
\begin{subfigure}[b]{0.49\linewidth}
\includegraphics[width=\textwidth]{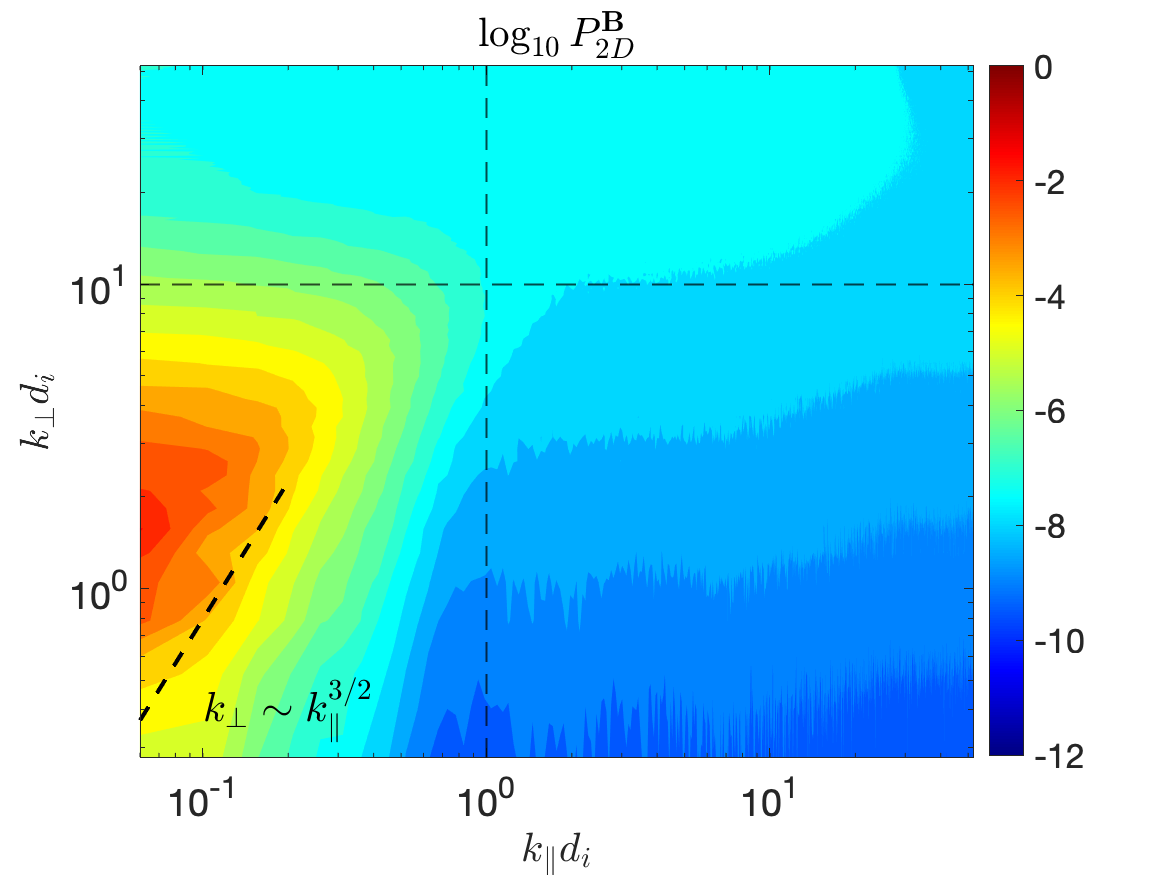}
\caption{$t=12/\omega_{pi}$}
\label{fig:2D12b}
\end{subfigure}
\begin{subfigure}[b]{0.49\linewidth}
\includegraphics[width=\textwidth]{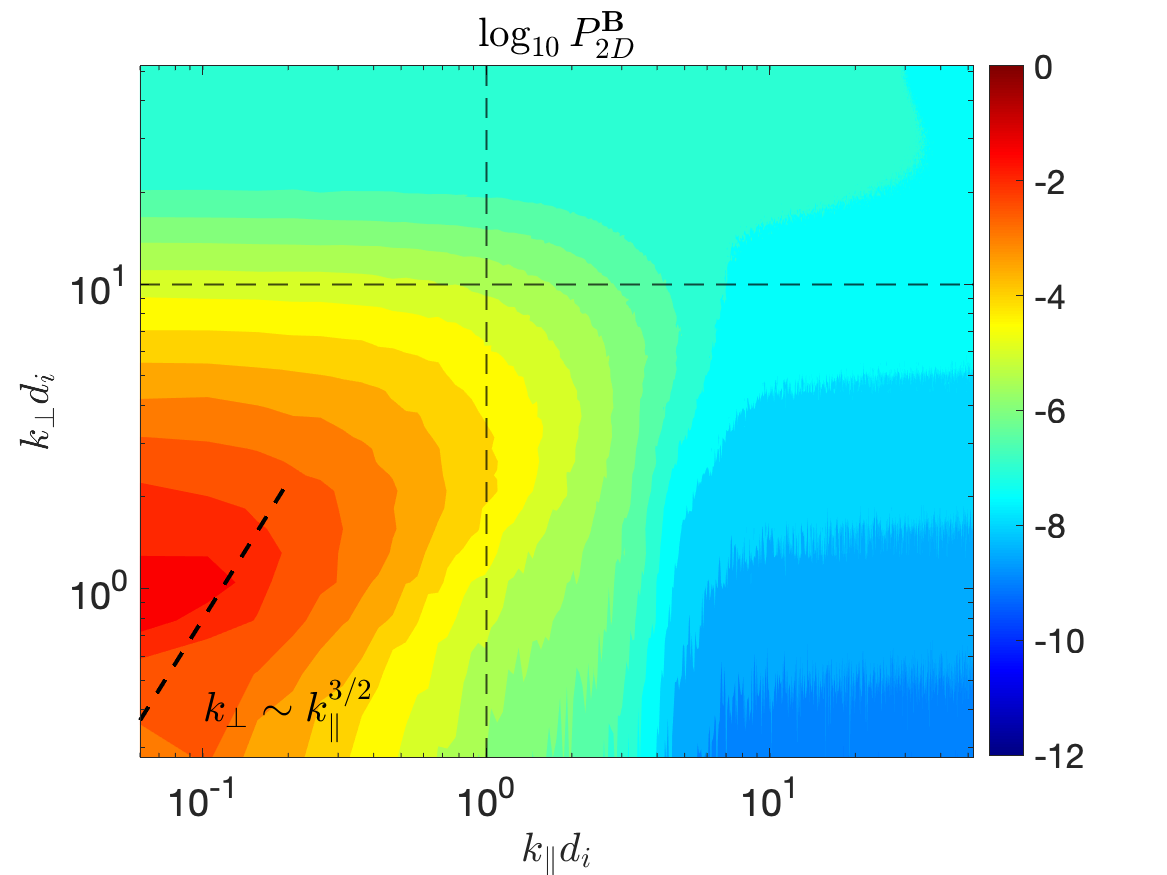}
\caption{$t=t_{R}$}
\label{fig:2D12c}
\end{subfigure}
\begin{subfigure}[b]{0.49\linewidth}
\includegraphics[width=\textwidth]{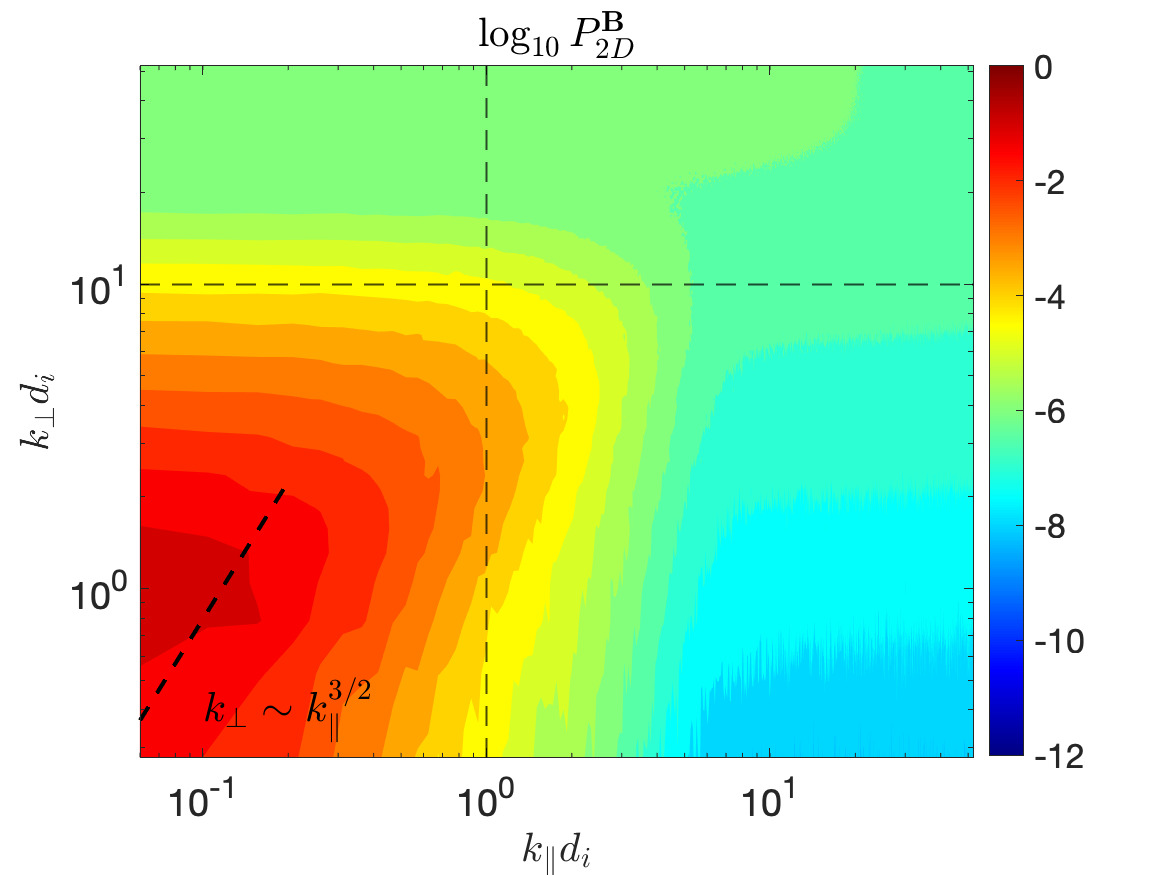}
\caption{$t=240/\omega_{pi}$}
\label{fig:2D12d}
\end{subfigure}
\caption{Isocontours of $\log_{10}P^{\mathbf{B}}_{2D}$ of the fluctuating magnetic field as a function of $k_{\parallel}$ and $k_{\perp}$ at different time-steps. The dashed lines provide a reference for the scaling of $k_{\perp}$ and $k_{\parallel}$. The horizontal (vertical) dashed line marks $k_{\perp}d_{e}=1$ ($k_{\parallel}d_{i}=1$). At $t=0$, the spectrum shows the modes of our initialisation and their Fourier harmonics. At $t=12/\omega_{pi}$, the cascade in the perpendicular direction (vertical axis) has proceeded beyond electron scales ($k_{\perp}d_{i}   \geq 10$). At $t=t_{R}$, although the perpendicular cascade has not proceeded significantly further, the cascade in the parallel direction (horizontal axis) has reached the kinetic range ($k_\parallel d_i\approx 1$) up to ion scales but not to electron scales. At $t=240/\omega_{pi}$ the distribution has not considerably changed compared to $t=t_{R}$}
\label{fig:2D12}
\end{figure}

\noindent Figure \ref{fig:2D12} shows the logarithm of the 2D reduced power spectral density of the magnetic-field fluctuations $P^{\mathbf{B}}_{2D}$ normalised to $\max{P^{\mathbf{B}}_{2D}}$ in the $k_{\parallel}$-$k_{\perp}$ plane at $t=0$ (panel (\textit{a})), $t=12/\omega_{pi}$ (panel (\textit{b})), $t=t_{R}$ (panel (\textit{c})) and $t=240/ \omega_{pi}$ (panel (\textit{d})). The horizontal dashed line marks $k_{\perp}d_{e}=1$ which corresponds to $k_{\perp}d_{i} = 10$ owing to our mass ratio of $m_i/m_e=100$. The vertical dashed line marks $k_{\parallel}d_{i}=1$. At $t=0$ the energy is entirely stored in the initial modes. At $t=12/\omega_{pi}$, the isocontours show that the energy has already cascaded to $k_{\perp} d_e >1$ whereas the parallel cascade has not yet reached the kinetic range. At $t=t_{R}$ the perpendicular cascade has not proceeded any further but the parallel energy transport reached $k_{\parallel} d_i >1$. At $t=240/\omega_{pi}$ the energy distribution has not considerably changed compared to the distribution at $t=120/\omega_{pi}$. For comparison with analytical predictions, we overplot the expected critical-balance scaling of $k_{\perp}\sim k_{\parallel}^{3/2}$ as a dashed line at small $k_{\perp}$. We note, however, that $P_{2D}^{\mathbf{B}}$ exhibits a broad distribution in $\mathbf{ k}$-space around this prediction. In order to explore the anisotropy of the cascade in more detail, we compute the perpendicular one-dimensional reduced power spectral density

\begin{align}
P^{\psi}_{1D_{\perp}}(k_{\perp}) = \int_{0}^{\infty} P^{\psi}_{2D}(k_{\perp},k_{\parallel})dk_{\parallel}, 
\label{eqn:1Dfour_per} 
\end{align}

\noindent and the parallel one-dimensional reduced  power spectral density

\begin{align}
P^{\psi}_{1D_{\parallel}}(k_{\parallel}) = \int_{0}^{\infty} P^{\psi}_{2D}(k_{\perp},k_{\parallel})dk_{\perp} 
\label{eqn:1Dfour_par}
\end{align}

\begin{figure}
\centering
\begin{subfigure}[b]{0.49\linewidth}
\includegraphics[width=\textwidth]{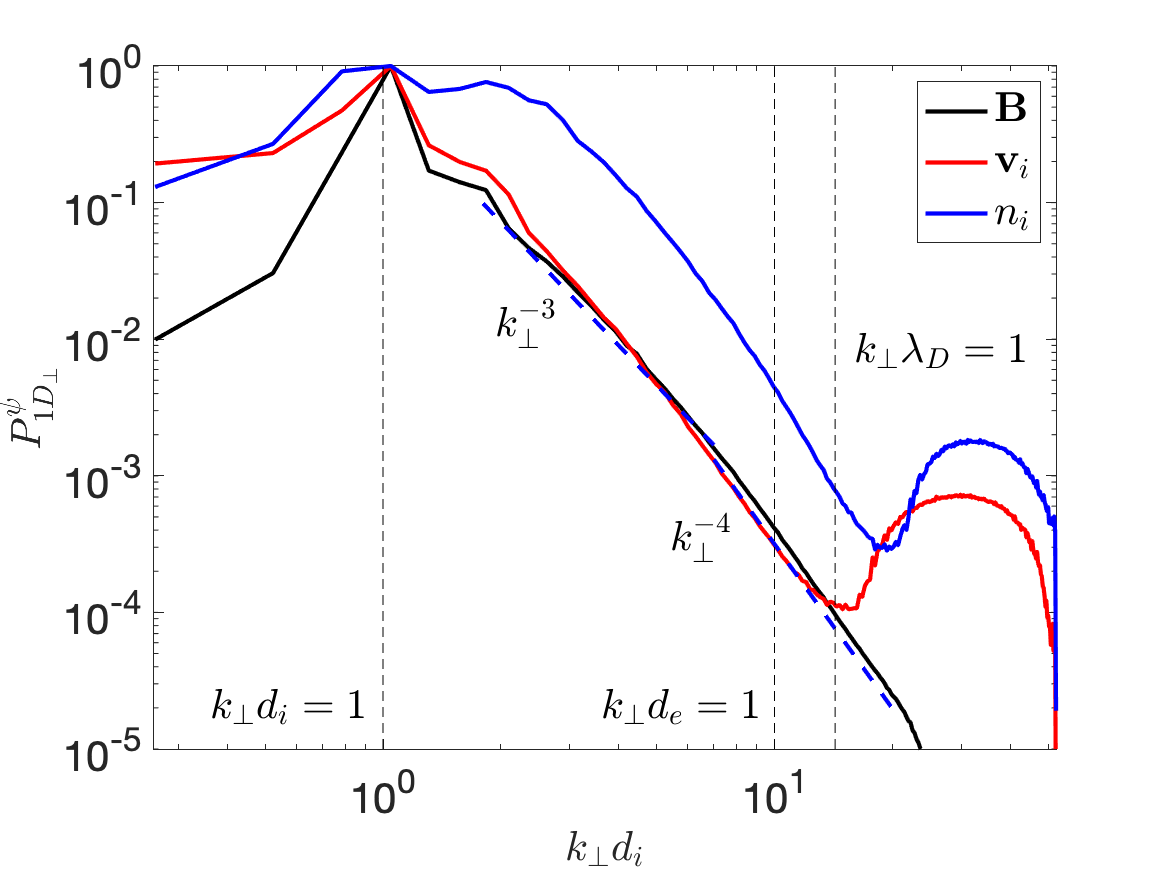}
\caption{}
\label{fig:1D12a}
\end{subfigure}
\begin{subfigure}[b]{0.49\linewidth}
\includegraphics[width=\textwidth]{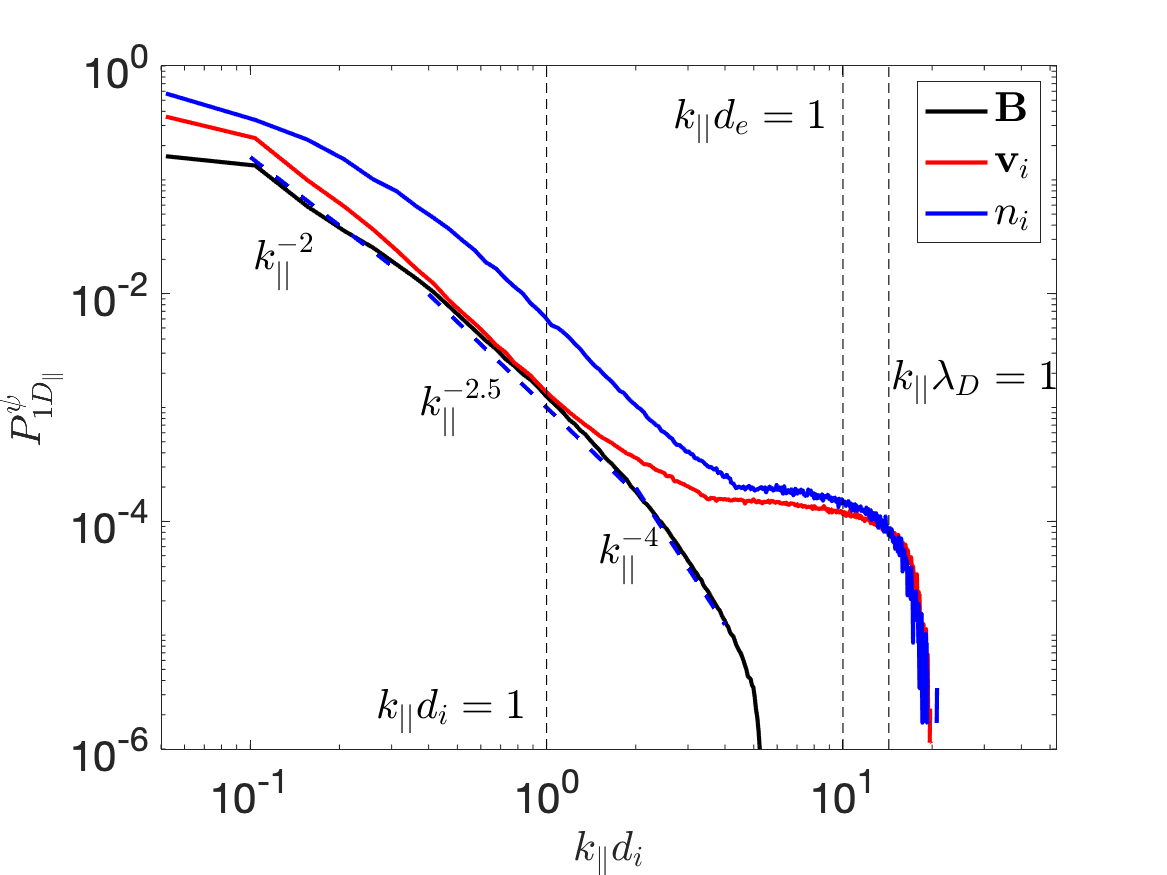}
\caption{}
\label{fig:1D12b}
\end{subfigure}
\caption{(\textit{a}) Perpendicular and (\textit{b}) parallel reduced one-dimensional power spectral densities $P^{B}_{1D_{\parallel, \perp}}$ (black), $P^{v_{i}}_{1D_{\parallel, \perp}}$ (red)  and $P^{n_{i}}_{1D_{\parallel, \perp}}$ (blue) at  $t=t_{R}$. The vertical dashed lines indicate $k_{\parallel, \perp}d_{i} = 1$, $k_{\parallel, \perp}d_{e} = 1$ and $k_{\parallel, \perp}\lambda_{D} = 1$.} 
\label{fig:1D12}
\end{figure}

\noindent of multiple fluctuating quantities $\psi$. Panel (\textit{a}) in Figure \ref{fig:1D12} shows the perpendicular one-dimensional reduced power spectral density of the magnetic-field fluctuations $P^{{B}}_{1D_{\perp}}$ (black line), of the ion velocity fluctuations $P^{{v}_{i}}_{1D_{\perp}}$ (red line) and of the ion density fluctuations $P^{{n}_{i}}_{1D_{\perp}}$ (blue line) at $t=t_{R}$. The vertical dashed lines mark $k_{\perp}d_{i}=1$, $k_{\perp}d_{e}=1$ and $k_{\perp}\lambda_{D}=1$. The enhancement in $P^{{v}_i}_{1D_{\perp}}$ at $k_{\perp}d_{i} = 17$ is an artefact created by Debye-length effects and the finite spatial resolution of the system. The scale of the initial waves in the perpendicular direction coincides with the transition point of the energy cascade from inertial to kinetic scales, i.e., $k_{\perp}d_{i}=1$. Therefore, our simulations do not describe the cascade at $k_{\perp}d_{i}\le 1$. During the first nonlinear time the system develops a broadband spectrum of perpendicular density fluctuations in the kinetic range. $P^{{B}}_{1D_{\perp}}$ and $P^{{v}_{i}}_{1D_{\perp}}$ exhibit similar spectral indices in part of the kinetic range between $k_{\perp}d_{i} \sim 3$ and $\sim 6$. Within the same interval, $P^{{n}_{i}}_{1D_{\perp}}$ follows a steeper spectrum. These features suggest the presence of both Alfvénic and compressive fluctuations, consistent with the presence of kinetic Alfvén waves. $P^{{B}}_{1D_{\perp}}$ in the interval $k_{\perp}d_{i} \sim 1.8$ to $\sim 7$, follows a power-law scaling with a spectral slope of $-3$. In the range between $k_{\perp}d_{i} \sim 7$ and $\sim 20$ the slope is slightly steeper with a power index of approximately $-4$ \footnote{We note that we observe a change in slope within a single decade in $k_{\perp}$. The interpretation of a change in slope over such a small range of scales must be interpreted with caution. Although it indicates a steepening in $P_{1D\perp}^{B}$ towards increasing $k_{\perp}$, the scale separation is insufficient to  apply Kolmogorov-like scaling arguments to these spectral slopes.}. Although we calculate the energy spectrum of the magnetic-field fluctuations using the global background magnetic field, these values are within the range of slope variability measured in the solar wind \citep{chen2010anisotropy,bruno2014spectral} as well as in hybrid simulations \citep{franci2018solar,gonzalez2019turbulent}.

\indent Panel (\textit{b}) in Figure \ref{fig:1D12} shows the parallel one-dimensional reduced power spectral density of the magnetic-field fluctuations $P^{{B}}_{1D_{\parallel}}$ (black line), ion velocity fluctuations $P^{{v}_{i}}_{1D_{\parallel}}$ (red line) and ion density fluctuations $P^{{n}_{i}}_{1D_{\parallel}}$ (blue line) at $t=t_{R}$. The vertical dashed lines mark $k_{\parallel}d_{i}=1$, $k_{\parallel}d_{e}=1$ and $k_{\parallel}\lambda_{D}=1$. At $k_{\parallel}d_{i} \le 1$, $P^{{B}}_{1D_{\parallel}}$ and $P^{{v}_{i}}_{1D_{\parallel}}$ follow a similar trend as expected for Alfv\'enic turbulence. The spectral slope for $P^{\tilde{B}}_{1D_{\parallel}}$ is close to $-2$ between $k_{\parallel}d_{i} \sim 0.1$ and $\sim 0.3$ which is in agreement with the magnetic-field power spectrum $k_{\parallel}^{-2}$ observed in the solar wind \citep{bavassano1989evidence, grappin1991alfvenic, wicks2010power, wicks2011anisotropy, chen2011anisotropy}. At smaller parallel scales, the spectrum steepens to $-2.5$ between $k_{\parallel}d_{i} \sim 0.4$ and $\sim 2$ and further towards $-4$ between $k_{\parallel}d_{i} \sim 2$ and $\sim 4$. Both the perpendicular and parallel spectral indices have values of -4. The equality of these exponents has been observed in 3D hybrid PIC simulations and has been suggested to be a consequence of the anisotropy being frozen at sub-proton scales \citep{franci2018solar,arzamasskiy2019hybrid,cerri2019kinetic,landi2019spectral}. Although we initialise the system with non-compressive waves, the simulation swiftly develops a cascade of density fluctuations which suggests that compressive modes form self-consistently in the energy cascade. The development of compressive fluctuations has been suggested to depend on the plasma parameters rather than the initial conditions \citep{cerri2017plasma}. The level of compressive fluctuations in our simulation is greater than observed in the solar wind \citep{chen2016recent}, but the reasons for the creation of such strong compressive fluctuations is unknown. At $k_{\parallel}d_{i} \approx 1.4$, the slope of $P^{{v}_{i}}_{1D_{\parallel}}$ separates from the slope of $P^{\tilde{B}}_{1D_{\parallel}}$ and approaches the slope of $P^{{n}_{i}}_{1D_{\parallel}}$. The flattening of $P^{{n}_{i}}_{1D_{\parallel}}$ at $k_{\parallel}d_{i} \approx 4$ is due to finite particle noise. 


\subsection{Reconnection sites}
\label{subsec:finding_sites}

In this section, we confirm that magnetic reconnection occurs in our simulation domain. Methods to find reconnection sites in 2D simulations are based on the identification of magnetic islands and their closest x-point within a current sheet \citep{wan2014dissipation,papini2019can}. However, the interaction of magnetic structures such as flux tubes, which are the 3D equivalent of 2D magnetic islands, is more complex than in the 2D case and magnetic reconnection does not happen at a single point but in an extended region \citep{daughton2011role,liu2013bifurcated,daughton2014computing}. In 2D and 3D theories of reconnection, strong current sheets are often associated with reconnection events as the key locations of energy dissipation. However, there are events in which the x-points are not placed exactly within the current-sheet \citep{priest1995three,Wan2014}. The presence of a strong guide magnetic field and asymmetries of the reconnection event can shift the position of the x-point and even preclude the reconnection event \citep{eastwood2010asymmetry,eastwood2013influence}. Moreover, proton temperature anisotropies in reconnection events can trigger kinetic instabilities, which then have a stabilising effect on the current sheet \citep{matteini2013signatures}.

\indent In our turbulent simulation setup, we expect that once the reconnection events occurred, most of them exhibit local asymmetries due to the turbulent nature of the domain. Moreover, the background magnetic field acts as a guide field in reconnecting flux ropes. Therefore, in order to capture all reconnection events in such a complex and asymmetric field geometry, we require a new method to determine reconnection sites in our 3D simulations. Strong gradients in at least one component of the magnetic field as well as magnetic null points are common features of both 2D and 3D reconnection events. Strong gradients directly related to the presence of current sheets according to Amp\`ere's law. The presence of magnetic null points is not a requirement for reconnection. In 2D reconnection, for instance, the presence of a guide field removes this requirement \citep{hesse2004role}. 3D reconnection, on the other hand, can take place in collapsing structures that form current sheets related to quasi-separator lines, which do not require magnetic null points \citep{pritchett2004three, pontin2011three}. Exhaust regions in which particles are accelerated to velocities near the Alfv\'en speed are another common feature. Magnetic reconnection not only accelerates particles but also increases their thermal energy. Hence, an enhancement in the population of heated particles is a further indicator of reconnection as long as it occurs near a region in which accelerated particles and magnetic field gradients are present. 

\indent During magnetic reconnection, the electric field is responsible for the energy exchange between particles and fields in the current sheet. The associated heating is quantified by $\mathbf{J} \cdot \mathbf{E}$ \citep{somov1985magnetic, ni2016heating}. We expect to find coherent regions in the simulation domain in which $\mathbf{J} \cdot \mathbf{E}$ is non-zero. According to 3D steady-state theories of magnetic reconnection \citep{hesse1988theoretical,priest2003nature,pontin2011three}, when a magnetic field line enters a diffusion region, the integral of the parallel electric field($E_{\parallel}=\mathbf{E}\cdot \mathbf{B}/| \mathbf{B}|$) along the magnetic field line within the diffusion region must be different from zero. Since a non-zero $E_{\parallel}$ can indicate the presence of non-vanishing diffusive terms in Ohm's law, we use the presence of non-zero $E_{\parallel}$ as a possible indicator for a diffusion region located within a finite volume.  Although $E_{\parallel}$ is not a good indicator in the absence of a guide magnetic field, we expect to find coherent regions in the simulation domain with non-zero $E_{\parallel}$. 

\indent In summary, we identify the following indicators that we consider essential for the presence of reconnection in a region of our simulation domain. We adopt a clustering detection method \citep{uritsky2010structures} based on the mean value of each quantity $\psi$, its rms value $\psi^{rms}$ and a threshold value $N_{th}$. Thus, we search the simulation domain for regions in which $\psi  \geq  \langle \psi \rangle +N_{th}(\psi)^{rms}$. Our indicators for magnetic reconnection are: 
\\
\begin{enumerate}[leftmargin=*,labelindent=16pt,label=C\arabic*]
\item \ Current-density structures,  $|\mathbf{J}|  \geq  \langle |\mathbf{J}| \rangle +N_{th}(|\mathbf{J}|)^{rms}$ \footnote{We note that given the ambiguity in the definition of current sheets when studying observational data, the indicator C1 can be defined as $\nabla \times \mathbf{B}$ instead of $|\mathbf{J}|$.};
\item \ Fast ions and electrons, $|\mathbf{v}_{i,e}|  \geq  \langle |\mathbf{v}_{i,e}| \rangle +N_{th}(|\mathbf{v}_{i,e}|)^{rms}$;
\item \ Heated particles,  $T_{i,e}  \geq  \langle T_{i,e} \rangle +N_{th}(T_{i,e})^{rms}$; 
\item \ Energy transfer between fields and particles, $| \mathbf{J} \cdot \mathbf{E} - \langle |\mathbf{J} \cdot \mathbf{E}| \rangle | \geq  N_{th}(|\mathbf{J} \cdot \mathbf{E}|)^{rms}$;
\item \ Non-zero parallel electric fields, $| E_{\parallel} - \langle |E_{\parallel}| \rangle | \geq  N_{th}(|E_{\parallel}|)^{rms}$.\\
\end{enumerate}

\indent To find the number of events satisfying these conditions, we use the first-neighbour volumetric method described in Section \ref{subsec:time_evolution}. We apply the algorithm to identify clusters of contiguous cells fulfilling each condition separately as well as combinations of them. Afterwards we apply a filter to remove all regions with an equivalent volume $V \leq 1d_{i}^{3}$, where $V$ is defined as the sum of the volumes of all contiguous cells associated with the cluster. This is motivated by the fact that we are mostly interested in events in which both ions and electrons experience reconnection. Therefore, we expect to find coherent regions with a size of at least $d_i$. We analyse two values for the threshold: $N_{th}=3$ and $N_{th}=4$. We present our results in table \ref{tab:conditions}, where C2$_{i}$ and C2$_{e}$ refer to the separate application of criterion C2 to ions and to electrons respectively.  The same definitions apply to C3. C4$_{+}$  and C4$_{-}$ refer to the application of condition C4 separated by cases in which $\mathbf{J}\cdot \mathbf{E} >0$ (+) and $\mathbf{J}\cdot \mathbf{E} <0$ (-). The same definitions apply to C5. As expected, a larger number of locations fulfil these conditions if the threshold is lower. Moreover, all events detected with $N_{th}=4$ are also detected when using $N_{th}=3$. There are no events that fulfil our condition C5. The reason for this result is that, although local regions fulfil C5, the size of contiguous volumes of cells fulfilling C5 are never greater than $1d_i^3$. We attribute this effect to particle noise, which has a strong effect on parallel electric fields in PIC simulations. If we reduce the threshold to $N_{th}=2$, the algorithm is also unable to define clusters of cells, because our method is based on intensity thresholds which perform well for quantities with heavy tail distributions. The distribution of $E_{\parallel}$ in our simulation is spread with $\langle |E_{\parallel}| \rangle = 2.0 \times 10^{-3} B_{0}c$ and standard deviation $(|E_{\parallel}|)_{std} = 1.5 \times 10^{-3} B_{0}c$. The same argument applies to $\mathbf{J} \cdot \mathbf{E}$. Despite detecting at least 17 regions fulfilling C4 with $N_{th}=4$, there are no regions that satisfy all conditions C1 through C4 within a volume greater than $1d_{i}^3$. However, if we reduce the equivalent volume threshold to $0.3d_{i}^3$, we find 6 regions that fulfil conditions C1 through C4. We mark the corresponding numbers with an asterisk in Table \ref{tab:conditions}.

\begin{table}
  \begin{center}
\def~{\hphantom{0}}
    \begin{tabular}{lcccccccccc}
    \hline
      $N_{th}$ & C1 & C2$_{e}$ & C2$_{i}$ & C3$_{e}$ & C3$_{i}$
      &  C4$_{+}$ & C4$_{-}$ & C5$_{+}$ & C5$_{-}$  \\[3pt]
       3   & 149 & 144 & 77 & 92 & 82 & 68 & 77 & 0 & 0\\
       4   & ~97 &~ 92 & 29 & 50 & 39 & 23 & 17 & 0 & 0\\[3pt]
  \end{tabular}
  \begin{tabular}{lcccccc}
  \hline
      $N_{th}$ &  C1 and C2$_{i,e}$ & C1 and C3$_{i,e}$ & C1 through C3$_{i,e}$ & C1 through C4$_{-}$ & C1 through C4$_{+}$  \\[3pt]
       3   & 34 & 55 & 24 & 3 & 3 \\
       4   &   9 & 27 &  6 & $6^{*}$ & $6^{*}$\\
       \hline
  \end{tabular}
  \caption{Number of events in our simulation domain at time $t=t_R$ fulfilling each condition }
  \label{tab:conditions}
  \end{center}
\end{table}

\begin{figure}
\centering
\begin{subfigure}[]{0.495\linewidth}
\includegraphics[width=1\linewidth]{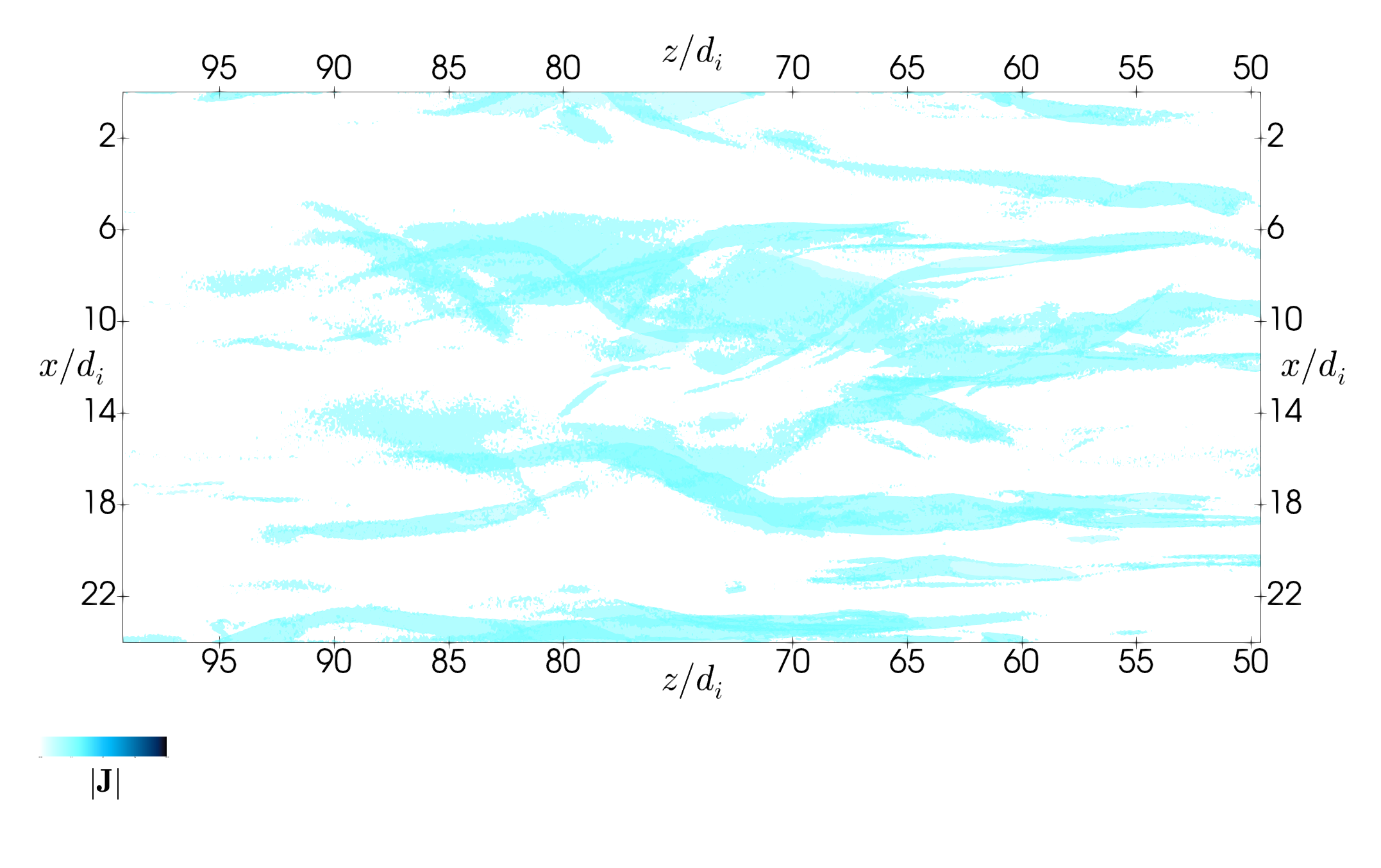}
\caption{Indicator C1}
\label{fig:rec_123a}
\end{subfigure}
\begin{subfigure}[]{0.495\linewidth}
\includegraphics[width=1\linewidth]{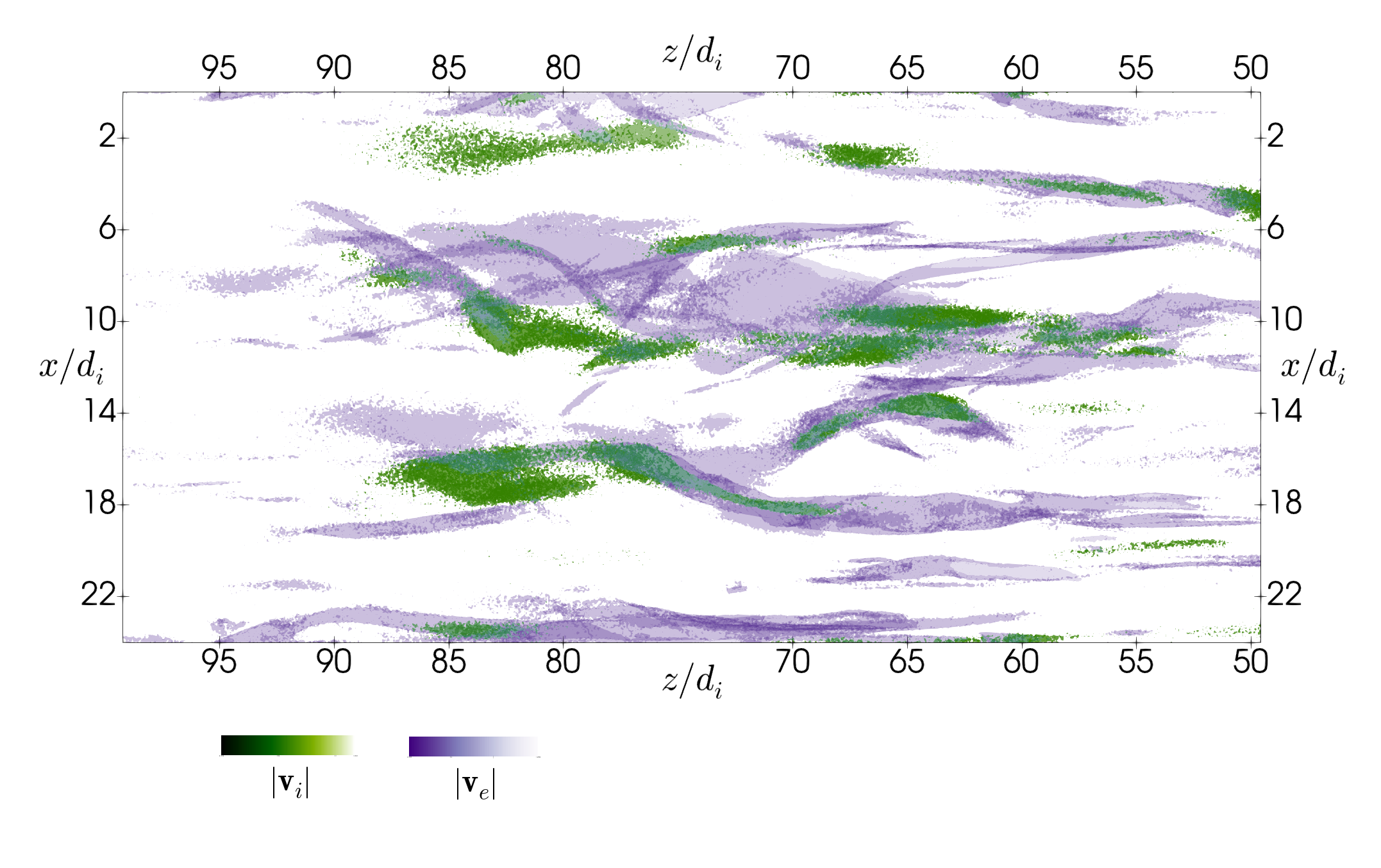}
\caption{Indicator C2}
\label{fig:rec_123b}
\end{subfigure}
\begin{subfigure}[]{0.495\linewidth}
\includegraphics[width=1\linewidth]{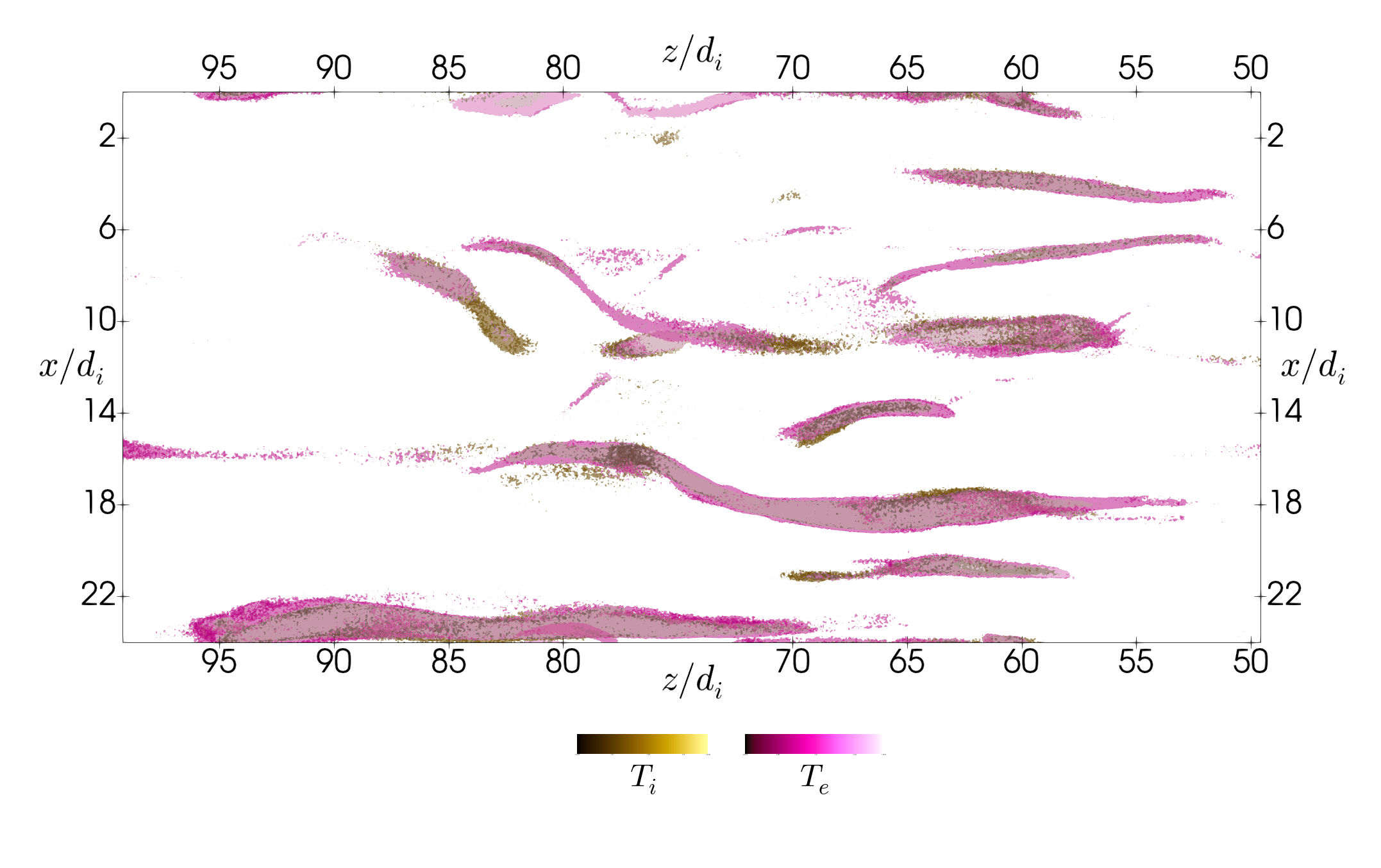}
\caption{Indicator C3}
\label{fig:rec_123c}
\end{subfigure}
\begin{subfigure}[]{0.495\linewidth}
\includegraphics[width=1\linewidth]{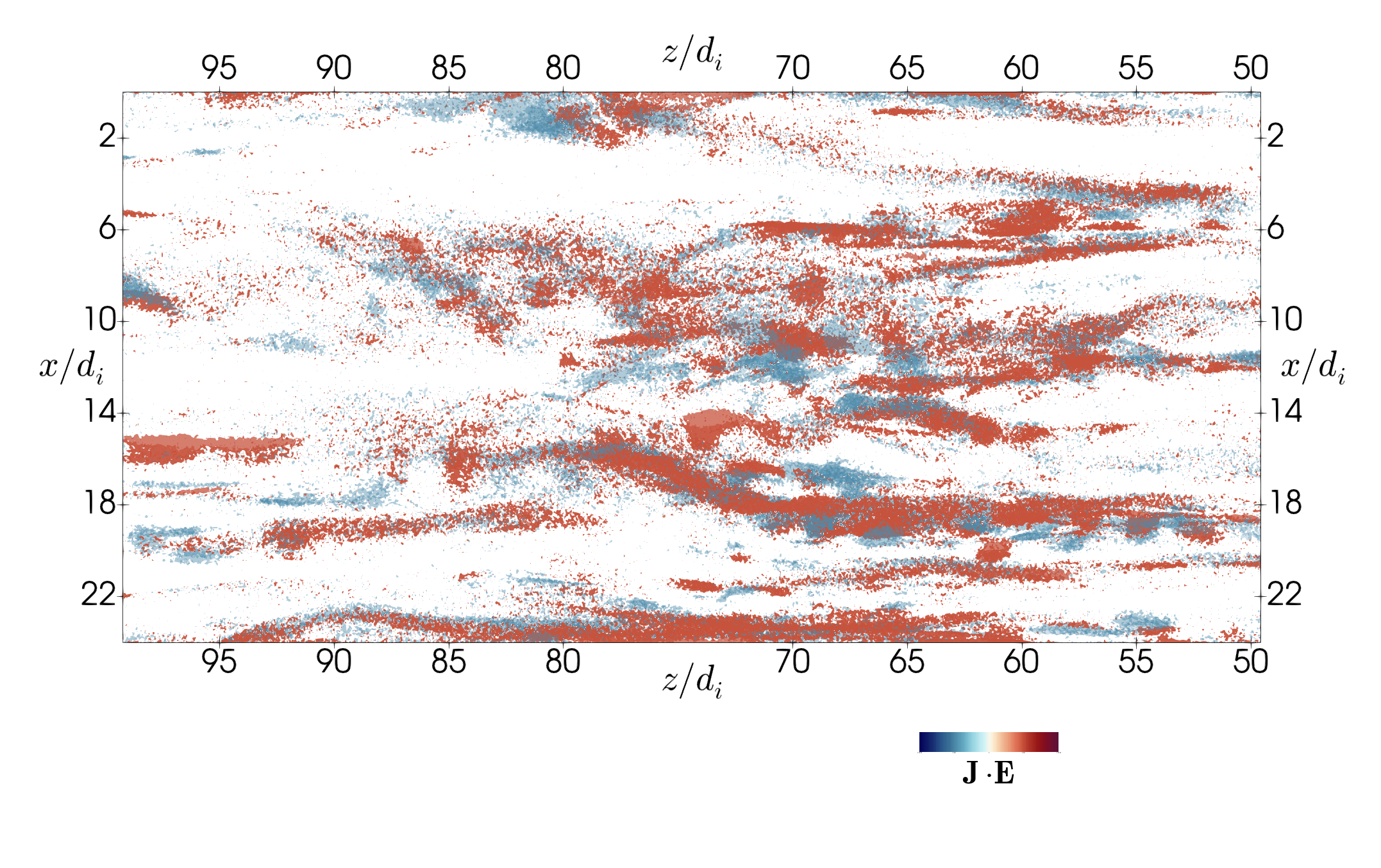}
\caption{Indicator C4}
\label{fig:rec_123d}
\end{subfigure}
\begin{subfigure}[]{0.495\linewidth}
\includegraphics[width=1\linewidth]{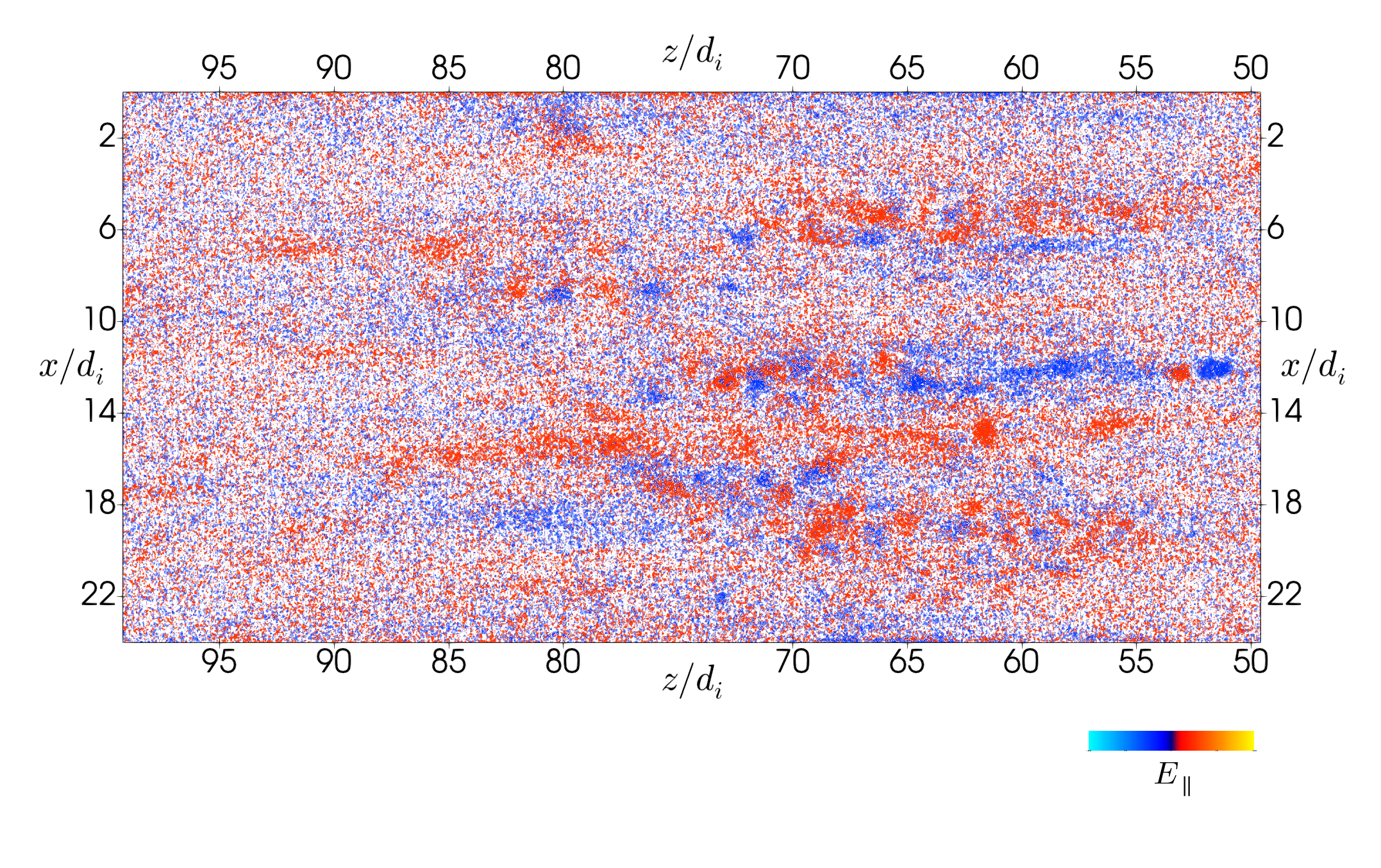}
\caption{Indicator C5}
\label{fig:rec_123e}
\end{subfigure}
\begin{subfigure}[]{0.495\linewidth}
\includegraphics[width=1\linewidth]{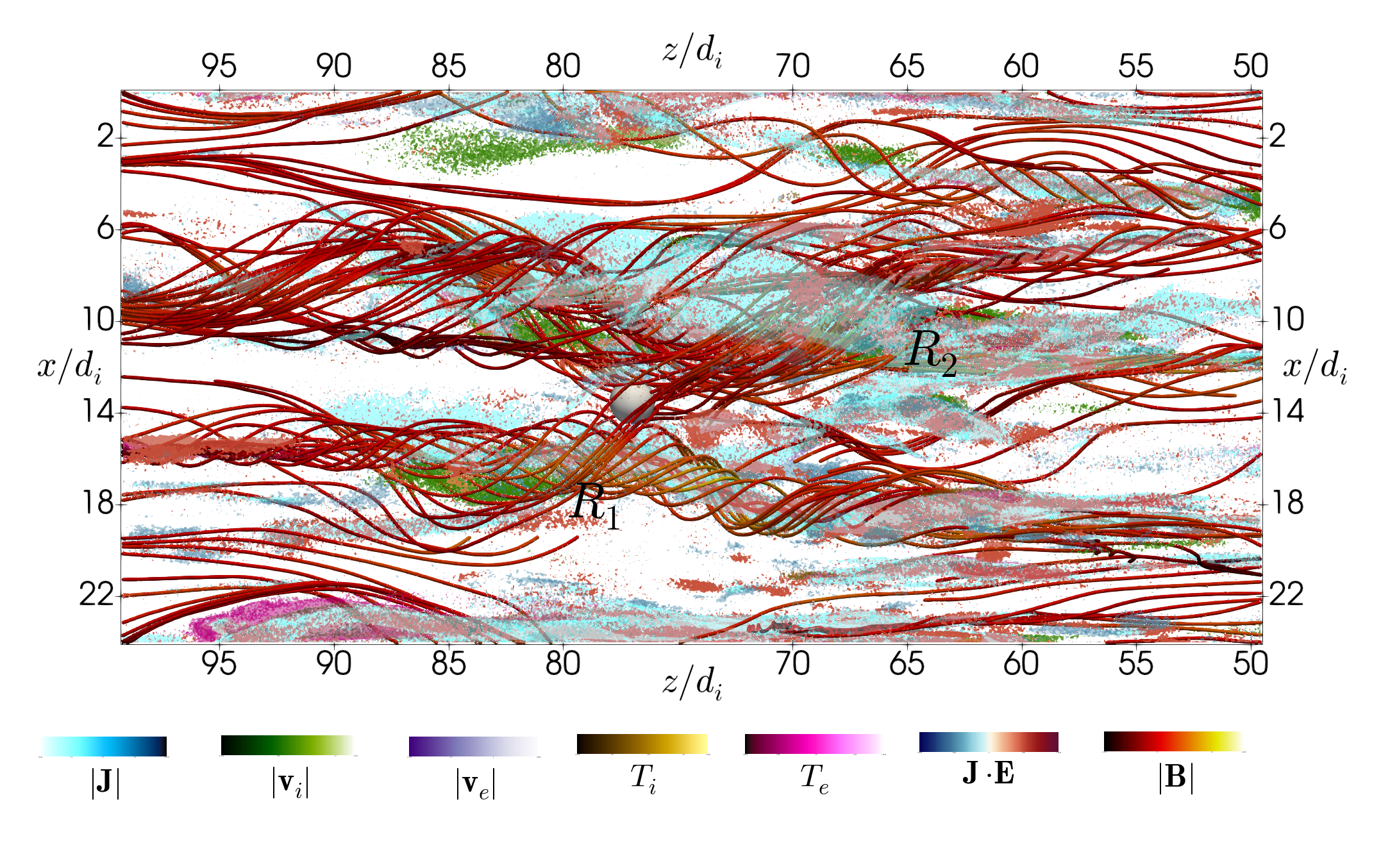}
\caption{Indicators C1 through C4}
\label{fig:rec_123f}
\end{subfigure}
\caption{Reconnection indicators projected onto a 2D cut in the $zx$-plane at $y=21d_{i}$. (\textit{a}) Indicator C1: Isosurfaces of $|\mathbf{J}|  =  \langle |\mathbf{J}| \rangle +3(|\mathbf{J}|)^{rms}$ (light blue). (\textit{b}) Indicator C2: Isosurfaces of $|\mathbf{v}_{i,e}|  =  \langle \mathbf{v}_{i,e} \rangle +3(\mathbf{v}_{i,e})^{rms}$ for ions (green) and for electrons (purple). (\textit{c}) Indicator C3: Isosurfaces of $T_{i,e}  =  \langle T_{i,e} \rangle +3(T_{i,e})^{rms}$ for ions (gold) and for electrons (pink). (\textit{d}) Indicator C4: Isosurfaces of $\mathbf{J} \cdot \mathbf{E} = \langle \mathbf{J} \cdot \mathbf{E} \rangle \pm 3(\mathbf{J} \cdot \mathbf{E})^{rms}$ for positive $\mathbf{J} \cdot \mathbf{E}$ (red) and negative $\mathbf{J} \cdot \mathbf{E}$ (blue). (\textit{e}) Indicator C5: Isosurfaces of $E_{\parallel} = \langle |E_{\parallel} |\rangle \pm 2(|E_{\parallel}|)^{rms}$ for positive $E_{\parallel}$ (orange) and negative $E_{\parallel}$ (blue). Panel (\textit{f}) shows, on top of the isosurfaces related to indicators C1 through C4, magnetic field lines colour-coded with $|\mathbf{B}|$. The magnetic field lines suggest the reconnection of a twisted flux rope with an adjacent flux rope. The white sphere of radius $1 d_i$ at $(z,x)=(77,13.5)d_{i}$ in panel (\textit{f}) is a reference point that marks the position of a reconnection site. In panel (f), we also indicate the regions $R_1$ and $R_2$ defined in the text. We provide a movie of the evolution of the magnetic field lines in the supplementary material.} 
\label{fig:rec_123}
\end{figure}

\indent In Figure \ref{fig:rec_123}, we visualise our indicators for magnetic reconnection. We use a 2D projection on the $zx$-plane of a part of our simulation domain, $50 d_{i} < L_{z} <100 d_{i}$. Panel (\textit{a}) shows the isosurfaces of $|\mathbf{J}|  =  \langle |\mathbf{J}| \rangle +3(|\mathbf{J}|)^{rms}$ (indicator C1) colour-coded in light blue. The selected structures mainly correspond to current filaments. Panel (\textit{b}) shows regions in which $|\mathbf{v}_{i}|  =  \langle \mathbf{v}_{i} \rangle +3(\mathbf{v}_{i})^{rms}$ (green) and $|\mathbf{v}_{e}|  =  \langle \mathbf{v}_{e} \rangle +3(\mathbf{v}_{e})^{rms}$ (purple), our indicator C2.  The locations of fast electrons according to C2 coincide with the locations of large currents according to C1, since the electrons are the main carriers of the electric current. This electron behaviour is consistent with observations in space plasma and reproduced in simulations \citep{phan2018electron}. We identify five structures in which accelerated ions coincide with our condition C1. Panel (\textit{c}) shows isosurfaces of $T_{i}  =  \langle T_{i} \rangle +3(T_{i})^{rms}$ (gold) and $T_{e}  =  \langle T_{e} \rangle +3(T_{e})^{rms}$ (pink), according to our indicator C3. Although the electric current is mostly carried by electrons, we find current structures that are not associated with high-temperature electrons and vice-versa.  The structures associated with heated electrons have mostly filamentary shapes. Panel (\textit{d}) shows the application of our indicator C4. The regions in which $\mathbf{J} \cdot \mathbf{E} = \langle \mathbf{J} \cdot \mathbf{E} \rangle \pm 3(\mathbf{J} \cdot \mathbf{E})^{rms}$ is positive (negative) are colour-coded in red (blue). There are large and diffuse clusters of positive and negative $\mathbf{J} \cdot \mathbf{E}$ between $z=55d_{i}$ and $z=85d_{i}$. We also locate filamentary structures of positive $\mathbf{J} \cdot \mathbf{E}$ which partially coincide with the regions fulfilling C3. Panel (\textit{e}) shows our indicator C5. The regions in which $E_{\parallel} = \langle |E_{\parallel} |\rangle \pm 2(|E_{\parallel}|)^{rms}$ is positive (negative) are colour-coded in orange (blue). The effect of particle noise on the electric field leads to difficulties in the determination of the associated clusters. Panel (\textit{f}) shows the combination of our indicators C1 through C4. We define two regions, $R_1$ and $R_2$, as the regions in which our indicators C1 through C4 are fulfilled. This suggests that magnetic reconnection is taking place in the vicinity of these regions. 

\indent To visualise the change of magnetic connectivity, we trace magnetic field lines in our simulation domain. The region of most intense $|\mathbf{B}|$ is co-located with $R_{1}$. The magnetic field lines suggest the reconnection of a twisted flux rope with an adjacent flux rope. The white sphere of radius $1 d_i$ at $(z,x)=(77,13.5)d_{i}$ is a reference region that marks the position at which the magnetic field lines associated with the flux ropes exchange connectivity. We provide a movie to support this claim in the supplementary material. The change of connectivity between the flux ropes lasts for $\sim 96/\omega_{pi} \sim 0.46 \tau_{nl}$, which is a long time compared to the time the turbulent cascade requires to develop. The long existence of connectivity exchange and of the current structure can be associated with the suppression of nonlinearities in the current sheet. In 2D geometries, the rate of magnetic-flux change between two magnetic islands, the so-called reconnection rate, is determined by the electric field at the x-point \citep{smith2004hall, servidio2011magnetic}. It can also be computed as the difference in the out of the plane component of the magnetic vector potential between the x-point and the o-point \citep{franci2017magnetic, papini2019can}. In 3D the reconnection rate can be computed integrating $E_{\parallel}$ along the magnetic field lines crossing the diffusion region \citep{schindler1988general,pontin2011three}. However, the complex structure of the field lines makes it unclear how to apply this method to our type of simulations \citep{liu2013bifurcated,daughton2014computing}. An extension of 2D methods that avoid the use of the electric field \citep{franci2017magnetic, papini2019can} to the 3D case requires the calculation of the vector potential which (a) is elaborate in 3D PIC simulations of the type used in this study and (b) impractical in the comparison with spacecraft data. 

\begin{figure}
\centering
\begin{subfigure}[]{0.493\linewidth}
\includegraphics[width=1\linewidth]{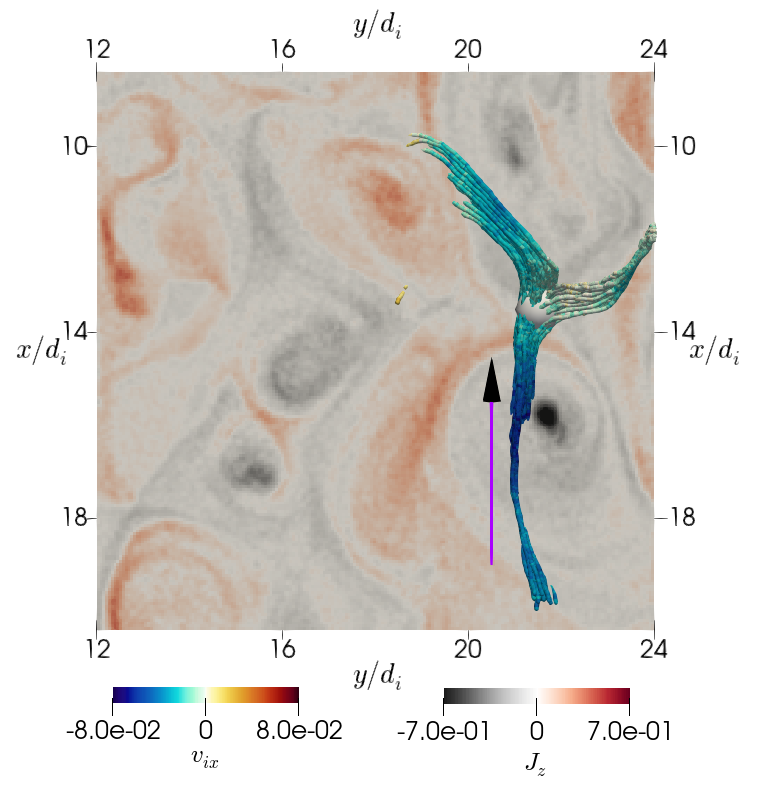}
\caption{}
\label{fig:rec_124a}
\end{subfigure}
\begin{subfigure}[]{0.493\linewidth}
\includegraphics[width=1\linewidth]{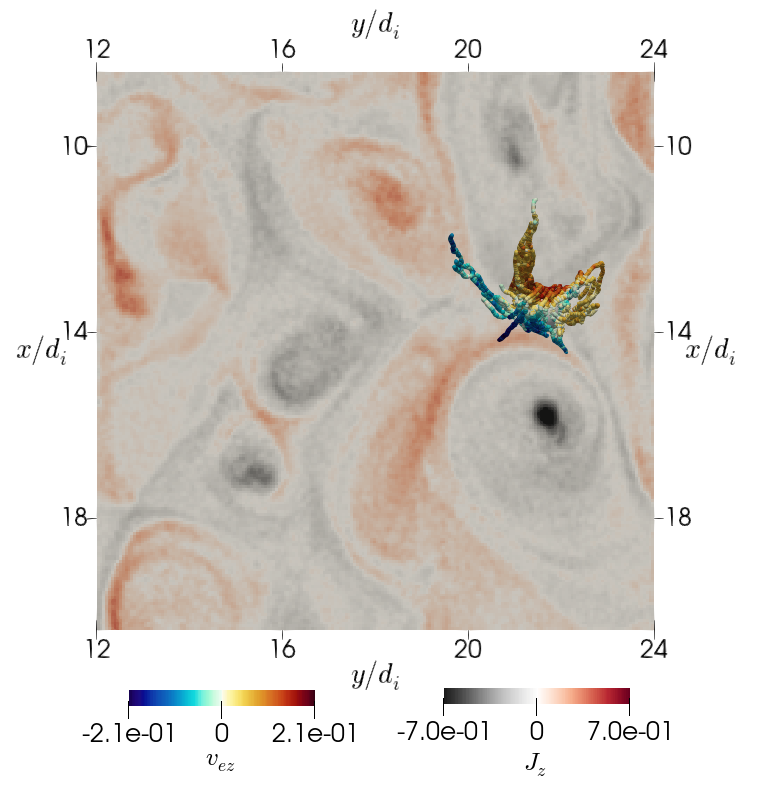}
\caption{}
\label{fig:rec_124b}
\end{subfigure}
\begin{subfigure}[]{0.493\linewidth}
\includegraphics[width=1\linewidth]{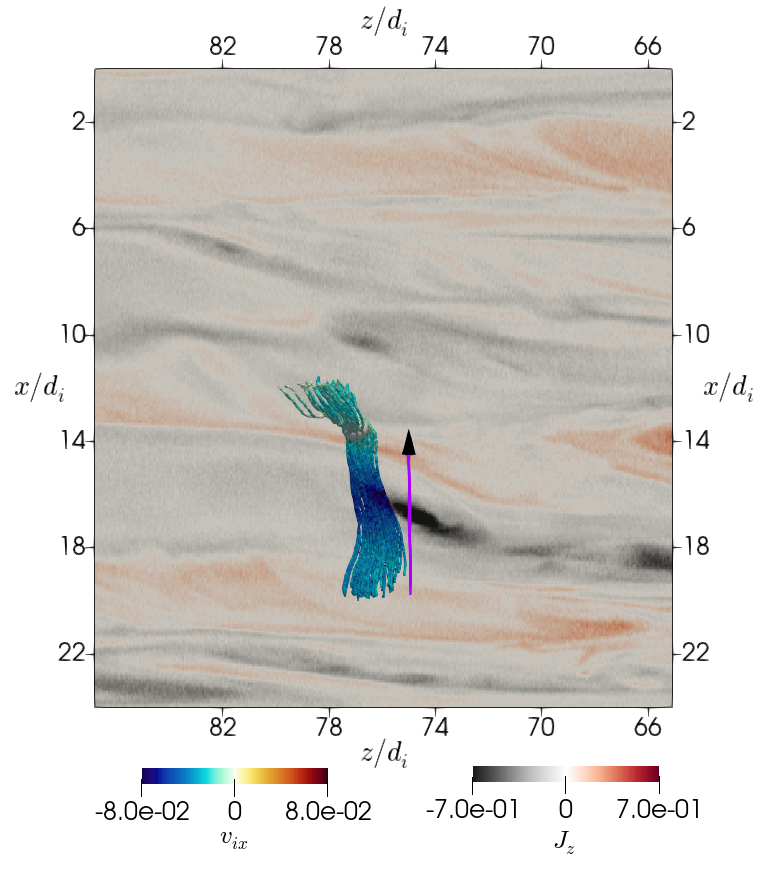}
\caption{}
\label{fig:rec_124c}
\end{subfigure}
\begin{subfigure}[]{0.493\linewidth}
\includegraphics[width=1\linewidth]{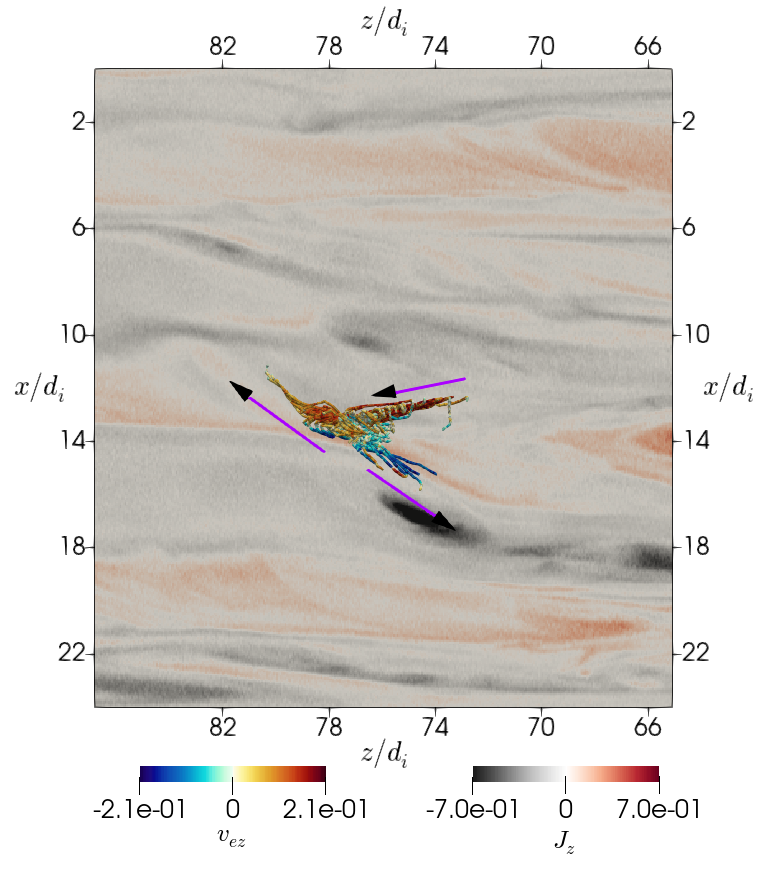}
\caption{}
\label{fig:rec_124d}
\end{subfigure}
\caption{Streamlines of the ion and electron bulk velocities over 2D cuts of the simulation plane showing $J_{z}$. (\textit{a}) and (\textit{b}) view over the $xy$-plane in which the $x$-direction points downward and the $y$-direction points towards the right-hand side. (\textit{c}) and (\textit{d}) over the $zx$-plane in which the $x$-direction points downward and the $z$-direction points towards the left-hand side. (\textit{a}) and (\textit{c}) show ion bulk velocity streamlines colour-coded with $v_{ix}$. (\textit{b}) and (\textit{d}) show electron velocity streamlines colour-coded with $v_{ez}$. The arrows indicates the direction of the ion bulk motion and of the electron bulk motion.}
\label{fig:rec_124}
\end{figure}

As the flux rope twists, it bends towards the region of changing magnetic connectivity, henceforth we refer to this region as the ``x-region''. During the flux-rope bending, plasma ions are accelerated towards the x-region. To illustrate this behaviour, we visualise the streamlines of the ion and electron bulk velocities that leave the reconnection region. Panel (\textit{a}) in Figure \ref{fig:rec_124} shows a view over an $xy$-plane cut of $J_{z}$. Gray colour represents negative values, red colour represents positive values, and white indicates a value of zero for $J_z$. The displayed streamlines of the ion bulk velocity emerge from the centre of the x-region. The streamlines are colour-coded with $v_{ix}$. The dark-blue segment near the dark-gray region indicates that the ions primarily move towards the reconnection site in the negative $x$-direction. As the ions approach the x-region, their speed decreases and their trajectories are deflected into the $y$-direction. The displayed streamlines maintain a coherent shape of width $\sim 2 d_{i}$ along the $z$-direction. Panel (\textit{c}) shows the same ion velocity streamlines but over an $zx$-plane cut of $J_{z}$. The region where ions have large $|v_{ix}|$ coincides with the core of the twisted flux rope in panel (\textit{f}) of Figure \ref{fig:rec_123} (black region) which suggests that they are accelerated by the bending of the flux rope. Considering the ion velocity streamlines, as indicative of the shape of the exhaust region associated with the x-region, the branch of the stream lines on the right-hand side in panel (\textit{a}) represents the reconnection exhaust of the event. It is three-dimensional and asymmetric. Likewise, the electron motion associated with the x-region is asymmetric. However, it differs considerably from the ion motion. Panel (\textit{b}) shows the electron velocity streamlines colour-coded with $v_{ez}$ in the same view as in panel (\textit{a}). These streamlines remain contained within a smaller region compared to the ion streamlines. They are mainly aligned with the $z$-direction. On the left-hand side of the reconnection site in panel (\textit{d}), the electron streamlines are directed along the $J_{z}$ structure as expected since the current is mostly (but not entirely) carried by electrons. In contrast, on the right-hand side of the reconnection site, the electrons move in directions towards and away from the reconnection site as is shown by the arrows. Considering the electron velocity streamlines, the electron exhaust is also asymmetric and three-dimensional but smaller than the ion exhaust. The diffusion region associated with the x-region of the reconnection event is likely to be the large structure of positive $J_{z}$ crossing the x-region in the $z$-direction in panel (\textit{c}). The shape of the electron streamlines suggests a diffusion region that resembles the distorted diffusion region observed in 3D Hall magnetic reconnection \citep{drake2008hall, yamada2014conversion}. 
\\
\indent In summary, our set of indicators suggests the presence of multiple reconnection sites in our simulation domain. Our automated identification based on our indicators allows for a detailed inspection of the magnetic-field connectivity of each event. Our method searches for clusters of cells fulfilling all conditions. This approach misses events in which ions and electrons are accelerated and heated in different locations near the reconnection site. If the event is large enough to affect both ions and electrons, we expect streams of accelerated particles for both species related to the reconnection event. Given the variability in the shape and size of these particle outflows, the volume threshold must be adjusted depending on the problem at hand in different simulation setups.

\subsection{1D trajectories across the reconnection region}
\label{subsec:1d_trajectories}

\begin{figure}
\centering
\begin{subfigure}[]{0.49\linewidth}
\includegraphics[width=1\linewidth]{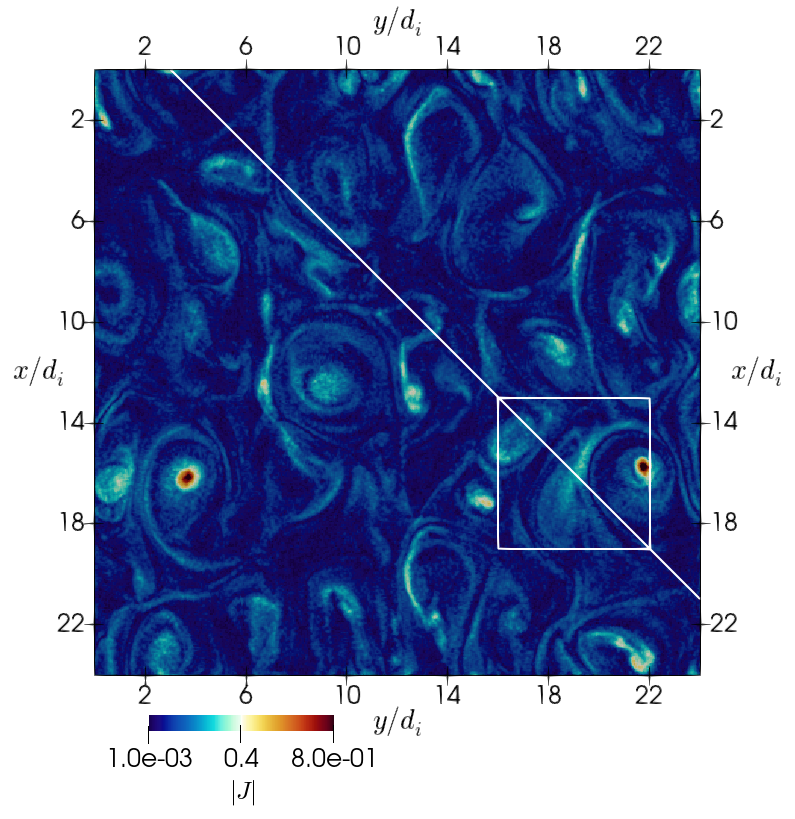}
\caption{}
\label{fig:rec_125a}
\end{subfigure}
\begin{subfigure}[]{0.49\linewidth}
\includegraphics[width=1\linewidth]{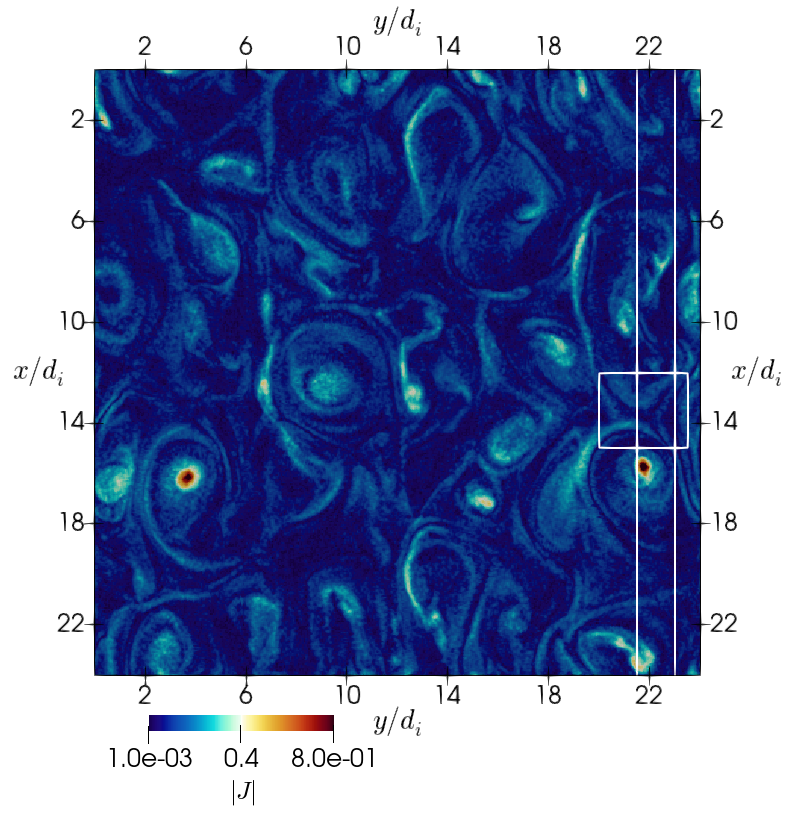}
\caption{}
\label{fig:rec_125b}
\end{subfigure}
\caption{Trajectories of an artificial spacecraft crossing our simulation domain. (\textit{a}) Trajectory \textit{T1}. The spacecraft moves from the top-left corner to the bottom-right corner. This trajectory crosses a region that we identify as a reconnection exhaust. The corresponding plasma and magnetic-field fluctuations are shown in panel (\textit{a}) of Figure \ref{fig:rec_1252}. (\textit{b}) Trajectories \textit{T2} and \textit{T3}. The corresponding plasma and magnetic-field fluctuations of \textit{T2} are shown in panel (\textit{d}) of Figure \ref{fig:rec_1252}.} 
\label{fig:rec_125}
\end{figure}

\begin{figure}
\centering
\begin{subfigure}[]{0.49\linewidth} 
(\textit{a})
\includegraphics[width=1\linewidth]{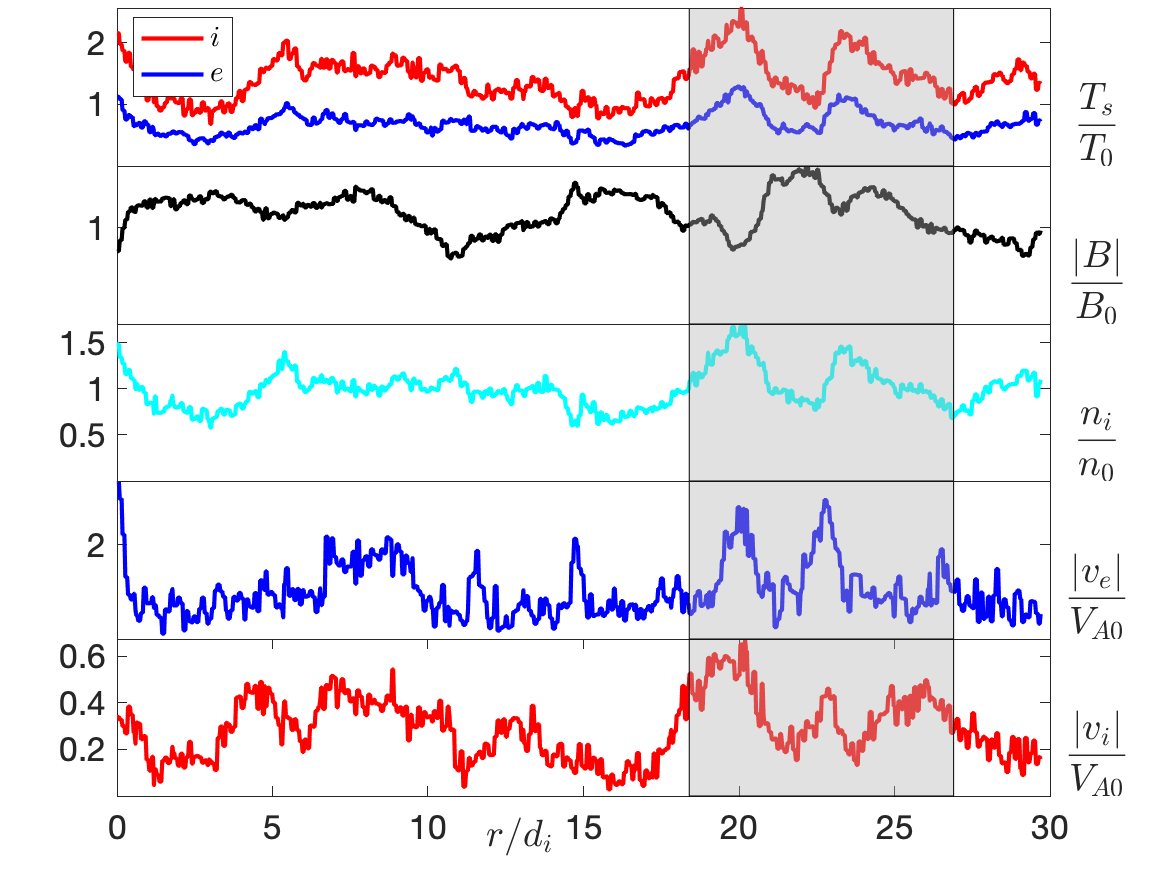}
\label{fig:rec_1252a}
\end{subfigure}
\begin{subfigure}[]{0.49\linewidth}
(\textit{d})
\includegraphics[width=1\linewidth]{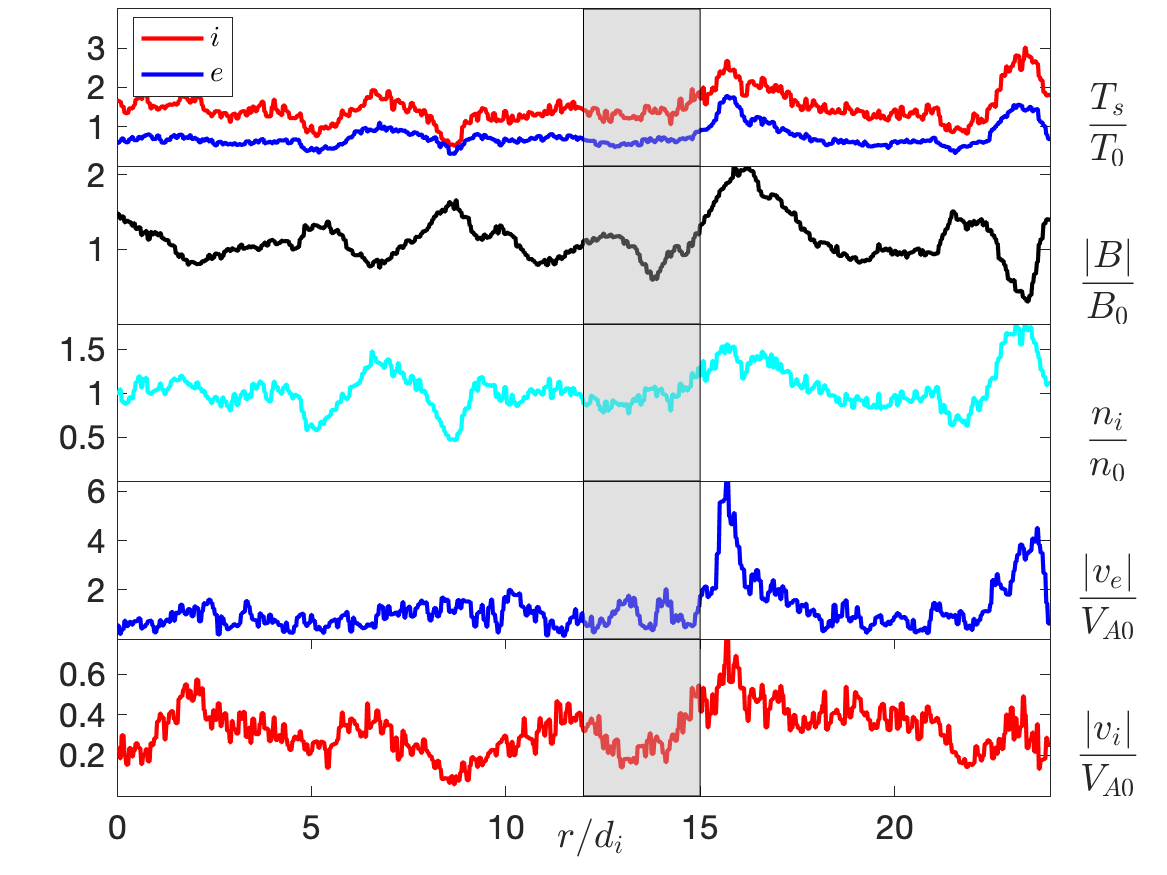}
\label{fig:rec_1252b}
\end{subfigure}
\begin{subfigure}[]{0.49\linewidth} 
(\textit{b})
\includegraphics[width=1\linewidth]{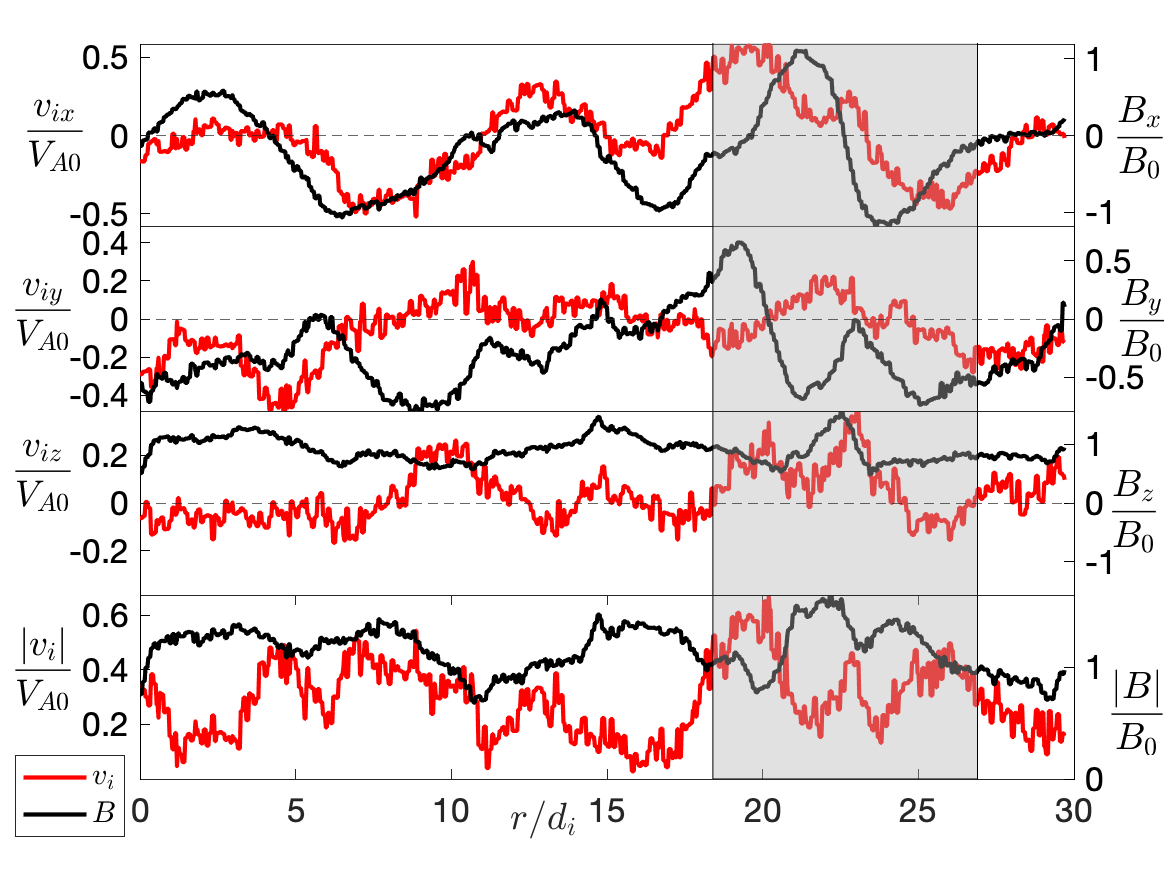}
\label{fig:rec_1252c}
\end{subfigure}
\begin{subfigure}[]{0.49\linewidth}
(\textit{e})
\includegraphics[width=1\linewidth]{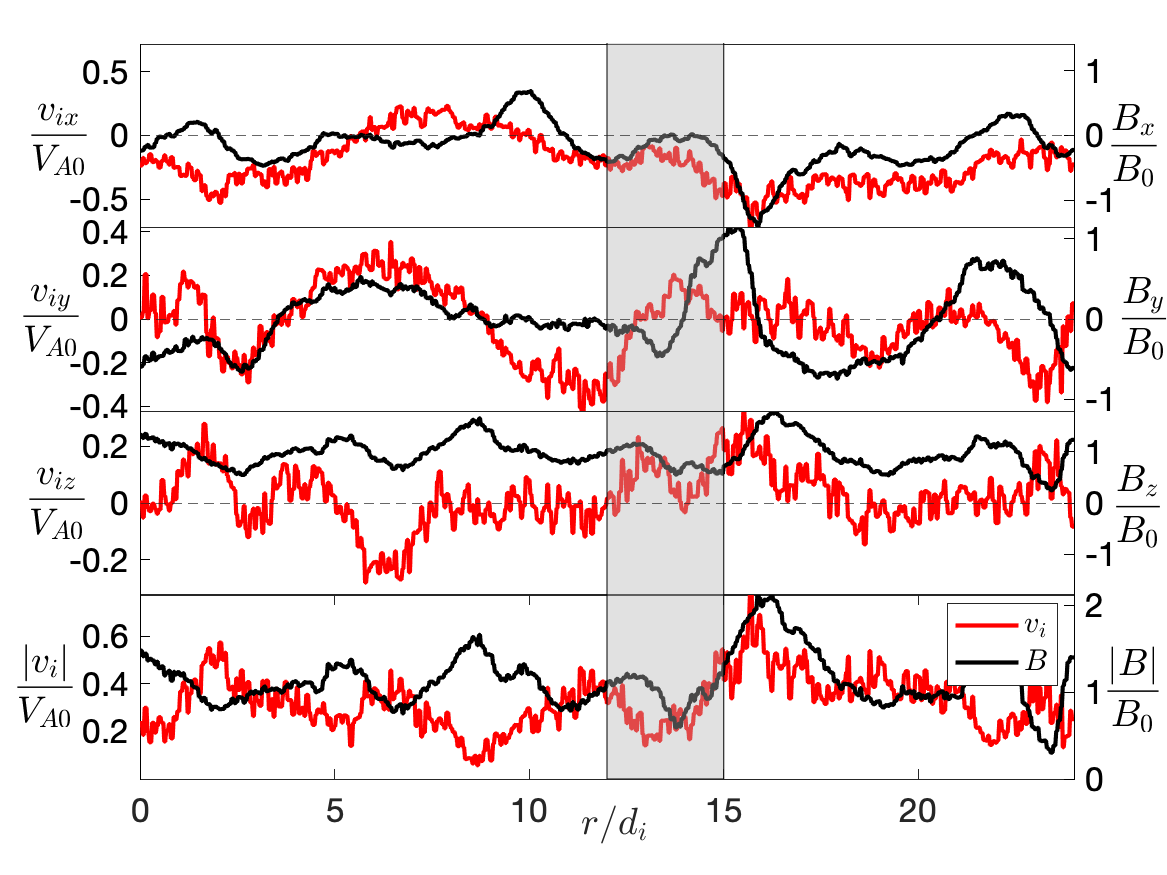}
\label{fig:rec_1252d}
\end{subfigure}
\begin{subfigure}[]{0.49\linewidth}
(\textit{c})
\includegraphics[width=1\linewidth]{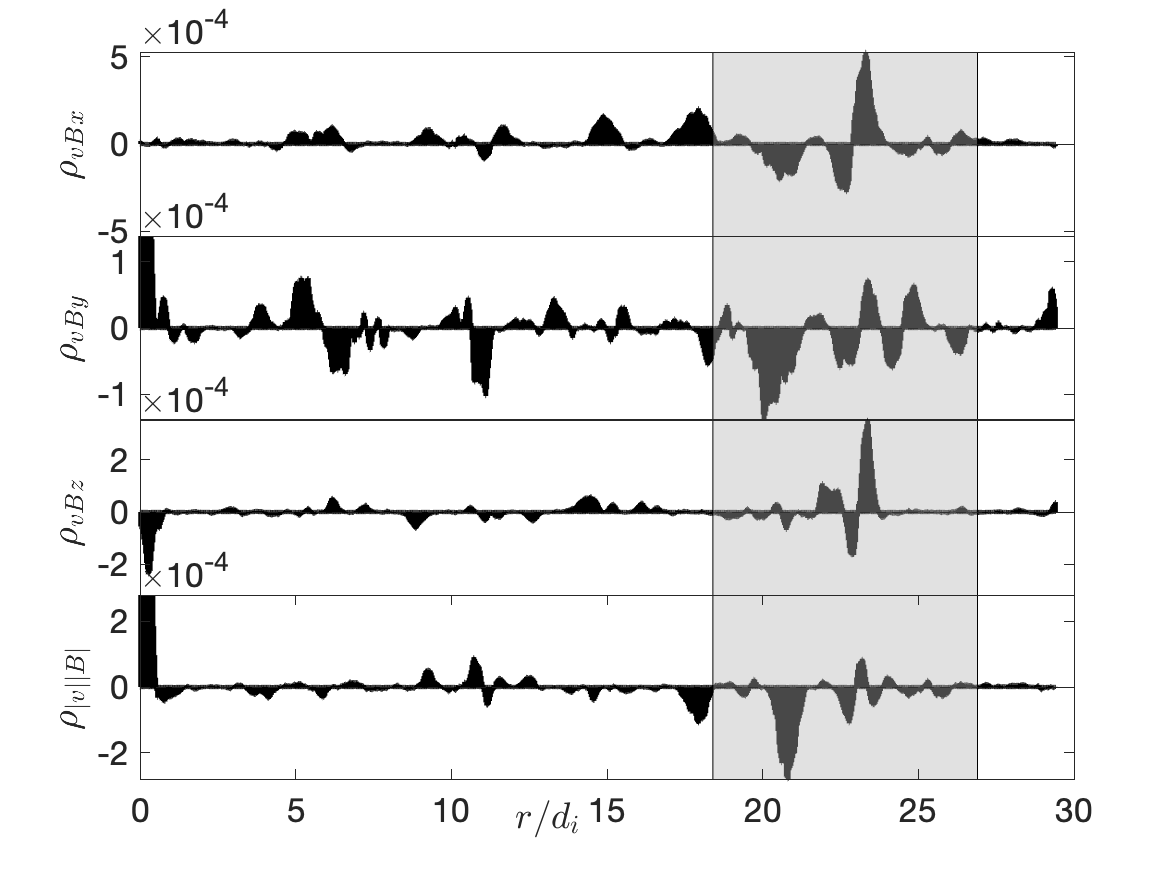}
\label{fig:rec_1252e}
\end{subfigure}
\begin{subfigure}[]{0.49\linewidth}
(\textit{f})
\includegraphics[width=1\linewidth]{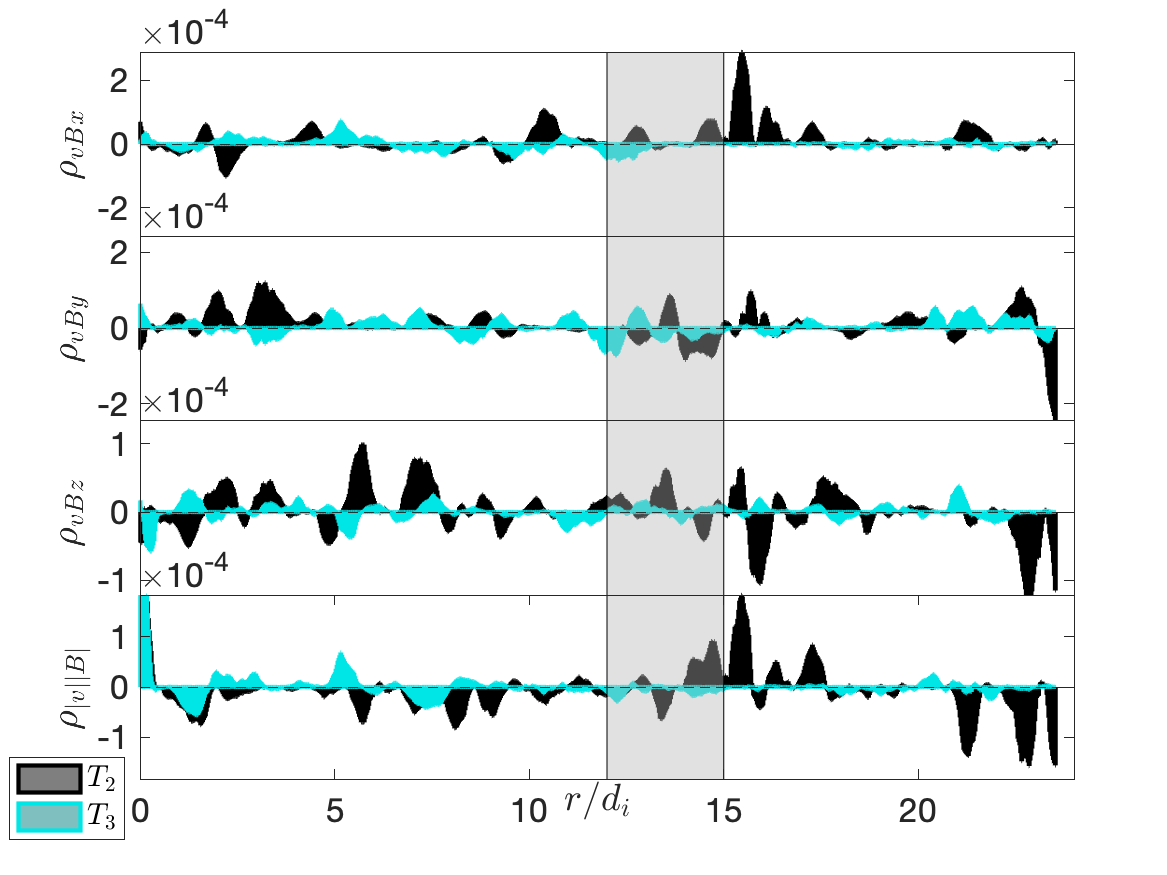}
\label{fig:rec_1252f}
\end{subfigure}
\caption{Left: plasma and magnetic-field fluctuations associated with our trajectory $\textit{T1}$. Right: plasma and magnetic-field fluctuations associated with our trajectory $\textit{T2}$. Panels (\textit{a}) and (\textit{d}) show the particle temperature $T_{i,e}$, magnetic field $B$, ion density $n_{i}$ and particle speed $v_{i,e}$ normalised as described in the text. The shaded areas mark the data recorded within the white squares in panel (\textit{a}) and (\textit{b}) of Figure \ref{fig:rec_125} respectively. Panels (\textit{b}) and (\textit{e}) show the components of the magnetic field (black) and ion velocity (red) for $\textit{T1}$ and $\textit{T2}$ respectively. Panels (\textit{c}) and (\textit{f}) show the derivative correlations of $\rho_{v_{i}B}$ and $\rho_{|v||B|}$ for trajectory $\textit{T1}$ and for trajectories $\textit{T2}$ and $\textit{T3}$ respectively.}
\label{fig:rec_1252}
\end{figure}

In-situ measurements of spacecraft typically record the plasma and magnetic-field fluctuations along the spacecraft trajectory. In order to compare such measurements with our 3D simulations, we ``fly'' an artificial spacecraft through our simulation box along three trajectories, \textit{T1}, \textit{T2} and \textit{T3}, and record the plasma and magnetic-field fluctuations along these trajectories. According to Taylor's hypothesis, we assume that the plasma structures are static as they are convected over the spacecraft with the average solar-wind bulk speed. The trajectories are taken within the $xy$-plane and are shown as the white lines in Figure \ref{fig:rec_125}. The trajectory \textit{T1}, shown in panel (\textit{a}) of Figure \ref{fig:rec_125}, passes close to the reconnection site when it crosses the white square although it does not carry the spacecraft right through the centre of the x-region. Panel (\textit{a}) of Figure \ref{fig:rec_1252} shows the plasma and magnetic-field fluctuations for our trajectory \textit{T1}. We normalise these quantities to their initial values at the beginning of the simulation. Thus, the ion and electron temperatures are normalised to $T_{0}$. The magnetic field and its components are normalised to the initial background magnetic field $B_{0}$. The ion density is normalised to the initial density $n_{0}$. The ion and electron velocities are normalised to the initial Alfvén speed $V_{A0}$. The shaded area in panel (\textit{a}) of Figure \ref{fig:rec_1252} represents the region delimited by the white square in panel (\textit{a}) of Figure \ref{fig:rec_125}. The ion and electron temperatures are positively correlated with each other as well as with the density across this trajectory. The magnetic-field and ion-density fluctuations exhibit mainly anti-correlation with each other across the trajectory. This correlation directly reflects the presence of slow-mode-like compressive fluctuations. The electron speed shows local peaks at $r \sim 11 d_{i}$ and $r \sim 15 d_{i}$ with no associated peaks in ion speed. This behaviour suggests the presence of local mechanisms that accelerate electrons only. This behaviour resembles electron-only reconnection events \citep{phan2018electron,stawarz2019properties,sharma2019transition,mallet2020onset}. However, our indicators show that both ions and electrons interact with this reconnection region.
\\
\indent When the artificial spacecraft trajectory \textit{T1} enters the region marked with the white square in panel (\textit{a}) of Figure \ref{fig:rec_125}, it encounters a coherent structure which exhibits enhancement in the ion and electron temperatures by a factor of about 1.5 to 2 compared to the background level at $r \sim 20 d_{i}$. At this position, the spacecraft observes a decrease in the magnetic field associated with an increase in the particle speed as well as an increase in the particle density. These are characteristic features associated with slow-mode-like fluctuations and shocks. Since in the trajectories shown in this section, the particle bulk speed is always less than the local magnetosonic speed, these events are not slow-mode shocks but rather fluctuations with a slow-mode-like polarisation. At $r \sim 22 d_{i}$, there is a slight enhancement in the electron speed which corresponds to the spike within the two large eddies seen in the white square in panel (\textit{a}) of Figure \ref{fig:rec_1252}. At $r \sim 23 d_{i}$, the spacecraft observes another slow-mode-polarised region which corresponds to the large structure in the middle of the square. According to the \citet{petschek1964magnetic} model of magnetic reconnection, the exhaust of particles is limited by a pair of slow-mode shocks. However, in recent studies of reconnection in the solar wind \citep{phan2006magnetic, phan2009prevalence, gosling2012magnetic}, the boundaries of reconnection exhausts often lack these features. Instead, exhausts are typically characterised through a rotation in the magnetic field along with a change in the sign of the correlation between the particle speed and the magnetic field \citep{gosling2012magnetic,phan2020parker}, consistent with our simulation results.
\indent Panel (\textit{b}) of Figure \ref{fig:rec_1252} shows from top to bottom $B_{x}$, $B_{y}$, $B_{z}$ and $|B|$ in black as well as $v_{ix}$, $v_{iy}$, $v_{iz}$ and $|v_{i}|$ in red for trajectory \textit{T1}. In the shaded area (ie. near the reconnection site), the velocity component $v_{ix}$ changes its sign between $r \sim 19 d_{i}$ and $r \sim 25 d_{i}$ while $\mathbf{B}$ undergoes a partial rotation. During the same interval, $v_{iy}$ shows little variation and $B_{iy}$ reverses its sign. Since the background magnetic field dominates $B_z$, the variations in the magnetic components $B_{x}$ and $B_{y}$ are more pronounced than the variations in $B_{z}$. As seen in the profile of $v_{iz}$, although ions are mostly stationary in the direction parallel to the background magnetic field, they are accelerated in the parallel direction near the slow-mode-like fluctuations. We note that the velocity spikes and magnetic-field drop-offs as seen in the $z$-component of $\mathbf{B}$ to a certain degree resemble the properties of the magnetic-field switchbacks observed in the solar wind \citep{kasper2019alfvenic,mcmanus2020cross}. Moreover, the blue regions in panel (\textit{a}) of Figure \ref{fig:simut34} suggest the possibility of magnetic reversals within the simulation domain. A comparison and further study is required to establish a potential correspondence between our simulation and observational data.  

\indent To visualise the correlation between the magnetic-field and velocity components we define the derivative correlation between the two variables $v_{j}$ and $B_{j}$ as

\begin{align}
    \rho_{vBj} = \frac{\Delta v_{j}}{\Delta r }\frac{\Delta B_{j}}{\Delta r } 
\end{align}

where $\Delta r$ is a distance increment, $\Delta v_{j} = v_{j}(r + \Delta r) - v_{j}(r)$ and $\Delta B_{j} = B_{j}(r + \Delta r) - B_{j}(r)$. We use $\Delta r=0.6d_{i}$ to reduce the effect of noise when calculating the derivative while keeping the spatial step small to cover small-scale fluctuations. Panel (\textit{c}) of Figure \ref{fig:rec_1252} shows from top to bottom $\rho_{v B x}$, $\rho_{v B y}$, $\rho_{v B x}$ and $\rho_{|v||B|}$ for trajectory \textit{T1}, where $\rho_{|v||B|}$ is defined accordingly with the magnitudes of $\mathbf{v}$ and $\mathbf{B}$. The $v_{ix}$ and $B_{x}$ components exhibit mostly positive correlation along the trajectory. However, there are two strong peaks of anti-correlation within the shaded area. Likewise, the $v_{iy}$ and $B_{y}$ components show more variability in the correlations from positive and negative derivative correlations within the shaded area than outside the area. This is due to the transit of the artificial spacecraft through the slow-mode-like fluctuations.  In particular, around $r = 23 d_{i}$, all three components present a change from anti-correlation to positive correlation. The presence of a pair of slow-mode-like fluctuations along with a magnetic-field rotation suggests that this region is indeed an exhaust region similar to those reported in previous observational studies in the solar wind \citep{gosling2012magnetic}.  
\\
\indent Trajectory \textit{T2} (the white line on the left in panel (\textit{b}) of Figure \ref{fig:rec_125}) carries the spacecraft right through the centre of the x-region. In panel (\textit{d}) of Figure \ref{fig:rec_1252}, at $r\sim 5 d_i$ and $r\sim 9 d_i$, the artificial spacecraft records particle temperature minima associated with density cavities as well as local peaks in the magnetic field. As the spacecraft moves towards the x-region, within the shaded region, the particle temperature remains approximately constant. There is a local minimum in the magnetic field which corresponds to the centre of the x-region at $r = 14 d_{i}$. On either side of the x-region, we find small enhancements in the electron speed. These peaks, in addition to the electron streams in panel (\textit{d}) of Figure \ref{fig:rec_124}, suggest the presence of electron-only streams in the vicinity of the x-region. The ion speed decreases as the spacecraft enters the x-region and increases as the spacecraft leaves the x-region. After leaving this region, the spacecraft encounters the highly twisted flux rope at $r = 16 d_{i}$ where it records an enhancement in all bulk quantities as well as in the magnetic field. The pair formed by the x-region  and the closest twisted flux rope resembles the known pairs of x-points and magnetic islands known from 2D models of reconnection. At the end of the trajectory, at $r\sim 23 d_i$, the spacecraft encounters a slow-mode-polarised structure which corresponds to the bright structure in the right-bottom-corner in panel (\textit{b}) of Figure \ref{fig:rec_125}. Panel (\textit{e}) of Figure \ref{fig:rec_1252} shows the components of the magnetic field and ion bulk velocity for trajectory \textit{T2}. From $r =10 d_{i}$ to $r =16 d_{i}$, $B_{x}$ changes polarity and from $r =8 d_{i}$ to $r =15 d_{i}$, $B_{y}$ undergoes a partial rotation. The change in the sign of $v_{iy}$ at the point where the spacecraft enters the shaded area and its value of approximately zero at the point where it leaves the shaded area in \textit{T2}, shows a local stream of particles leaving the region along the $y$-direction. This corresponds to the right-hand side branch of the ion streamline velocity in panel (\textit{a}) of Figure \ref{fig:rec_124}. At  $r =13 d_{i}$, $v_{iz}$  presents a mild peak corresponding to a weak current sheet. Entering the shaded area and up to $r \sim 19 d_{i}$, $v_{ix}$ is negative along \textit{T2} consistent with the stream of ions described in Figure \ref{fig:rec_124}.
\\
\indent 
Trajectory \textit{T3} is parallel to trajectory \textit{T2}, and the separation of these trajectories is $1.5d_{i}$. Along trajectory \textit{T3} $B_{z}$ and $B_{y}$ as well as $v_{iz}$ and $v_{iy}$ follow approximately similar behaviours (not shown here).
However, the local variations along \textit{T2} are more pronounced as this trajectory crosses through the centres of multiple structures. Panel (\textit{f}) shows the derivative correlation of the magnetic field and velocity components for trajectories \textit{T2} (black) and \textit{T3} (cyan). Trajectory \textit{T2} shows stronger positive and negative correlations in all components due to the transit through the structures. For the $x$-component, the peak of positive correlation corresponds to the transit through the flux rope which is associated with particle acceleration. 



\section{Discussion and Conclusions}
\label{sec:conclusions}

We simulate plasma turbulence created by the collision of counter-propagating Alfv\'en waves with a wavevector anisotropy consistent with the GS95 theory of critical balance at the small-scale end of the inertial range. Our initial waves have wavenumbers near the spectral breakpoint from the inertial to the kinetic range of turbulence. This choice allows us to set up the system with Alfvén waves and let the system develop kinetic and compressive fluctuations in the kinetic range self-consistently and with an anisotropy reminiscent of the solar wind, with the aim of developing reconnection features consistent with solar-wind turbulence. The use of a strong anisotropy in the initial waves allows the system to undergo nonlinear interactions and to create flux ropes during the first nonlinear time, which is in agreement with earlier simulation work \citep{grovselj2018fully}. Our initial anisotropic setup reduces the simulation time that a fully 3D PIC simulation of turbulence without this imposed anisotropy would require in order to develop reconnection as a product of anisotropic turbulence. 

\indent The nonlinear interaction of the anisotropic waves self-consistently creates Alfv\'enic turbulence and generates magnetic-field and current-density structures such as current filaments and current sheets as part of the turbulent cascade \citep{howes2013alfven1,howes2015dynamical,howes2016dynamical}. The initial scaling between $L_{\parallel}$ and $L_{D}$, for the magnetic structures, is $L_{\parallel} \sim L_{D}^{2/3}$. At $t=t_{R}$, the magnetic structures satisfying $V > d_{i}^{3}$ maintain an anisotropy consistent with the initial conditions and follow a $L_{\parallel} \sim L_{D}^{0.7}$ scaling. Although theoretical predictions including those based on intermittency \citep{boldyrev2012spectrum, boldyrev2019role}, kinetic simulations \citep{cerri2017kinetic, cerri2019kinetic} and observations in the solar wind \citep{wang2020observational} suggest the scaling $L_{\parallel} \sim L_{D}^{2/3}$ at sub-proton scales, our analysis of structures with $V \leq d_{i}^{3}$ is more consistent with an isotropic scaling $L_{\parallel} \sim L_{D}$ which has been observed in hybrid simulations \citep{franci2018solar, arzamasskiy2019hybrid, landi2019spectral}. The change of anisotropy over time (Figure \ref{fig:shapes}) is also observed in the evolution of the 2D reduced power spectral density (Figure \ref{fig:2D12}). The anisotropy initially decreases due to the change in the mean value of the distribution of cross section diameters and of the elongation of the magnetic structures. 

\indent The spectral index of the corresponding perpendicular 1D power spectrum of the magnetic-field fluctuations in the kinetic range varies between $-3$ and $-4$. Meanwhile, the spectral index of the parallel power spectrum of the magnetic-field fluctuations varies from $-2$ in the interval $0.1\lesssim k_{\parallel} d_i \lesssim 0.3$ to $-4$ at subproton scales. These results show that the simulation develops an anisotropic turbulent cascade and the associated 3D structures predicted to contribute to reconnection as a dissipation mechanism for turbulence.

\indent The critical-balance theory of Alfvénic turbulence has been tested using gyrokinetic simulations \citep{howes2008kinetic,tenbarge2012evidence} and 3D PIC simulations \citep{grovselj2018fully}. The evolution and morphology of 3D reconnection events, starting from a Harris current-sheet configuration, have been studied at kinetic scales  \citep{hesse2001particle, pritchett2001kinetic, wiegelmann2001evolution, lapenta2006kinetic, vapirev2013formation, liu2013bifurcated, munoz2018kinetic, lapenta2020local}, as has been the effect of turbulence on the development of reconnection events \citep{daughton2014computing, lapenta2015secondary, pucci2017properties, papini2019fast}. However, little attention has been given to the occurrence of small-scale reconnection as a product of the turbulent cascade in a fully 3D geometry . Our study contributes to the understanding, identification and geometry of these reconnection events. \\
\indent We establish a set of indicators to find regions in which magnetic reconnection takes place in 3D PIC simulations consistent with magnetic reconnection theories. These indicators are based on the presence of current-sheet structures (C1), fast particles (C2), heated particles (C3), and diffusion regions marked by energy transfer between fields and particles (C4) and non-zero parallel electric fields (C5). Since our method is based on thresholds for the bulk quantities, the selected regions correspond to high-intensity structures. Our method uses fast ions as an indicator (C2). Thus, this method does not identify all reconnection events, especially not those related to electron-only reconnection. In a follow-up study, it is worthwhile to investigate the role of the threshold level for the identification of reconnection sites and the relaxation of ion-based conditions to enable the identification of electron-only reconnection events. Our method is applicable as a first approach in the exploration of reconnection events in large 3D PIC simulations in which the handling of the kinetic particle information is computationally expensive due to the large number of particles. 

\indent We identify three regions that fulfil our set of indicators C1 through C4 for $N_{th}=3$ and have an equivalent volume larger than $1d_{i}^{3}$. We also illustrate the working of our method in a subset of our simulation domain. We inspect the time evolution of the magnetic field lines and observe the change of connectivity between a highly twisted flux rope and a less twisted flux rope. We find a good agreement between the geometry of the flux ropes formed by turbulence in our simulation with the flux ropes formed by the turbulent disruption of a Harris current-sheet \citep{daughton2011role}. We observe the occurrence of a complex reconnection event in which the region of changing connectivity (x-region) has a volume of $\sim 12.5 d_{i}^{3}$. This event dissipates turbulent fluctuations in current structures of order a few $d_{i}$ which are smaller than the smallest events recently observed in the solar wind \citep{phan2020parker} and different from the events observed in space which are mostly very large interface regions between plasmas \citep{phan2006magnetic,gosling2007observations}. The occurrence of electron-only reconnection \citep{phan2018electron,stawarz2019properties} and electron-scale turbulent fluctuations suggests that events as the one we describe take place in the solar wind.

\indent Although there is good agreement between studies using the Harris configuration and solar wind observations \citep{mistry2016observations}, our event is considerably more complex than the idealised steady and non-turbulent Harris current-sheet configuration often invoked to study magnetic reconnection. The shape of our reconnection region is asymmetric and the regions in which particle heating and acceleration occur are mostly associated with current filaments rather than current sheets. This suggests that the twist of the flux ropes plays a crucial role for the particle heating in our simulation. In addition, this finding supports the notion that reconnection events occur in the solar wind through small-scale flux ropes \citep{crooker1996heliospheric, moldwin2000small}.  
\\
\indent We trace 1D artificial-spacecraft trajectories across the simulation domain to study the fluctuations in the bulk quantities $n_{i}$, $\mathbf{v}_{i,e}$, $T_{i,e}$ and $\mathbf{B}$. These samplings may facilitate direct comparisons between our simulations and spacecraft observations in the solar wind. Our trajectories \textit{T1} and \textit{T3} pass near the identified reconnection region, and our trajectory \textit{T2} crosses through the centre of the x-region. We observe the presence of slow-mode-polarised fluctuations as anti-correlated fluctuations in $n_{i}$ and $|B|$, rotations in the magnetic field and changes in the sign of the correlation between the magnetic field and the ion velocity consistent with reconnection exhausts observed in the solar wind \citep{gosling2012magnetic}. Our artificial-spacecraft trajectory \textit{T2} (panel (\textit{d}) in Figure \ref{fig:rec_1252}) shows that an enhancement in all bulk quantities, which may be associated with a reconnecting flux rope. Moreover, this trajectory suggests that the encounter of a magnetic minimum followed by an enhancement in all bulk quantities may be associated with the encounter of an x-region and a flux rope. Such a pair x-region/flux-rope corresponds to the traditional pair x-point/o-point in 2D models of reconnection. It would be worthwhile to compare our simulated spacecraft trajectories with spacecraft observations of small-scale reconnection events and reconnection exhausts in the solar wind. The instrumentation onboard Solar Orbiter and Parker Solar Probe has the appropriate time resolution for such a comparison.

\indent In our reconnection event ions and electrons behave differently as shown in Figure \ref{fig:rec_124}. Both ions and electrons move towards and away from the x-region but in different directions. Our trajectories in the vicinity of the reconnection event suggest that the slow-mode-like features associated with the partial rotation in the magnetic field and the change in the $\mathbf{v}_i$-$\mathbf{B}$ correlation are also present in these spontaneously created small-scale events. 
\\
\indent The finite number of particles per cell has an important effect on the determination of coherent regions of strong $E_{\parallel}$, our indicator C5. Therefore, C5 is not a good indicator when the number of particles per cell is $\lesssim 100$. Although 2D studies of turbulence, magnetic reconnection \citep{franci2020modeling} and plasma instabilities \citep{hellinger2018electron} are able to use considerably larger numbers of particles per cell ($\sim 1000$), our work requires the third dimension in order to model the turbulence and the complex reconnection geometry more appropriately \citep{howes2015inherently, lazarian20203d}. Nonetheless, the increasing computational power of high-performance-computing facilities will allow us to perform increasingly more accurate 3D PIC simulations and to test all of our indicators over a wider range of parameters. Before these methods become computationally viable, divergence-cleaning of the electric field \citep{jacobs2009implicit} is a possible route to reduce the effect of particle noise.
\\
\indent Our data set possibly includes further reconnection sites that can be studied in more detail in the future. In this project, we use bulk quantities to study the reconnection events. In our future work, it is worthwhile to study the changes in the particle distribution functions as a result of the identified small-scale reconnection events. Such a more detailed study of the associated particle kinetics will allow us to understand the energy exchange between fields and particles and the details of the energy dissipation through small-scale reconnection events in the solar wind. 
\\
\\
\indent Acknowledgements: J.A.A.R. is supported by the European Space Agency's Networking/Partnering Initiative (NPI) programme and the Colombian programme Pasaporte a la Ciencia, Foco Sociedad - Reto 3 (Educación de calidad desde la ciencia, la tecnología y la innovación (CTel)),3933061, ICETEX. D.V. is supported by STFC Ernest Rutherford Fellowship ST/P003826/1. D.V., R.T.W., C.J.O., and G.N. are supported by STFC Consolidated Grant ST/S000240/1. K.G. is supported by NSF grant AGS-1460190. This work was performed using the DiRAC Data Intensive service at Leicester, operated by the University of Leicester IT Services, which forms part of the STFC DiRAC HPC Facility (www.dirac.ac.uk). The equipment was funded by BEIS capital funding via STFC Capital Grants ST/K000373/1 and ST/R002363/1 and STFC DiRAC Operations Grant ST/R001014/1. DiRAC is part of the National e-Infrastructure.
\indent The analysed simulation data are available under https://zenodo.org/record/4313310.
\indent The authors would like to thank the three anonymous reviewers whose thoughtful comments have led to significant improvements of our manuscript.


\appendix

\section{Initial conditions of the simulation}
\label{app:initial}

We initialise our simulation with eight anisotropic low-frequency counter-propagating Alfv\'en waves within a box of volume $L_{x} \times L_{y} \times L_{z}$. We take the background magnetic field to be along the $z$-axis, $\mathbf{B}_{0}=B_{0} \hat{z}$, and set up our fluctuations with wavevectors following the theory of critical balance by GS95. According to GS95, turbulence is isotropic at the large-scale end of the inertial range and develops an anisotropic cascade of energy with respect to the local magnetic field. The anisotropic cascade of energy is associated with a wavevector anisotropy $k_{\parallel} \propto \left( |\mathbf{k}_{\perp}| \right)^{\gamma}$, where $k_{\parallel}$ and $k_{\perp}$ are the wavevector components in the directions parallel and perpendicular with respect to the local background magnetic field. The index $\gamma$ is a power index that is approximately constant in each wavevector range of the turbulent power spectrum. For the inertial range, $\gamma=2/3$. Since the fluctuations are isotropic at the large-scale end of the inertial range, we express the relation between $k_{\perp}$ and $k_{\parallel}$ as 

\begin{align}
    k_{\parallel}d_{i} = C \left( k_{\perp} d_{i} \right)^{2/3},
\label{eqn:critB}
\end{align}
 
\noindent where $C$ is a constant which is chosen so that $k_{\parallel}=k_{\perp}$ at the large-scale end of the inertial range, which we set up as $k_{\perp}d_{i}=10^{-4}$ consistent with observations \citep{wicks2010power,chen2012three}. We define $k_{m,\perp} = \sqrt{k_{m,x}^{2} +k_{m,y}^{2}}$ where the index $m$ refers to the mode of the wave. Since we use periodic boundary conditions in our simulation, we adjust the wavelengths of our initial modes $\lambda_{m,i}$ so that $L_i$ is an integer multiple of $\lambda_{m,i}$. Then, the wavevector components are

\begin{align}
    k_{m,x}=m\frac{2\pi}{L_{x}}; \quad k_{m,y}=m\frac{2\pi}{L_{y}}; \quad \text{and} \quad k_{m,z}=m\frac{2\pi}{L_{z}}.
\end{align}

Since we use only $m = 1$, we drop the index $m$ for simplicity. Each wave satisfies the Alfvénic polarisation relation

\begin{align}
    \frac{\delta \mathbf{u}_{s,\alpha}}{V_{A0}} = (-1)^{\alpha}  \frac{\delta \mathbf{B}_{\alpha}}{B_0}, 
    \label{eqn:alfpol}
\end{align}

\noindent where  $V_{A0}=B_{0}/\sqrt{\mu_{0}n_{i}m_{i}}$ is the Alfvén speed, $n_{i}$ is the ion density and $m_{i}$ is the ion mass. $\mathbf{u}_{s,\alpha}$ is the bulk velocity of the species $s$. The index $\alpha=1,...,8  $ refers to each wave. The four waves with odd $\alpha$ travel along the $z$-direction and the other four in the opposite direction. The amplitude $\mathbf{A}_{\alpha}$ of the perturbation $\delta \mathbf{B}_{\alpha}$ of each wave is perpendicular to both the background magnetic field $\mathbf{B}_{0}$ and to the wave's wavevector $\mathbf{k}_{\alpha}$. Thus, we write the components of the wavevector as      

\begin{align}
    k_{\alpha,x}=k_{\alpha,\perp}\cos{\phi_{\alpha}}
\end{align}

\noindent and

\begin{align}
 k_{\alpha,y}=k_{\alpha,\perp}\sin{\phi_{\alpha}}, 
\end{align}

\noindent where $\phi_\alpha$ is the azimuthal angle of $ \mathbf{k}_{\alpha,\perp}$. The waves propagating in the $+z-$direction have $\phi_{\alpha}=0, \pi, \pi/4$ and $5\pi/4$ whereas the waves propagating in the $-z-$direction have $\phi_{\alpha}=\pi/2, 3\pi/2, 3\pi/4$ and $7\pi/4$. This distribution of azimuthal angles produces a quasi-gyrotropic distribution of fluctuations in the plane perpendicular to the background magnetic field while keeping the initial magnetic field divergence-free. The components of the fluctuating fields for each wave are given by

\begin{align}
    \delta B_{\alpha,x} = - |\mathbf{A}_{\alpha}| \cos(k_{\alpha,x}x+k_{\alpha,y}y + (-1)^{\alpha + 1}k_{\alpha,z}z + \psi_{\alpha}) \sin\phi_\alpha
\end{align}

\noindent and 

\begin{align}
    \delta B_{\alpha,y} = |\mathbf{A}_{\alpha}| \cos(k_{\alpha,x}x+k_{\alpha,y}y + (-1)^{\alpha + 1}k_{\alpha,z}z + \psi_{\alpha})\cos\phi_\alpha.
\end{align}

\noindent where $\psi_{\alpha}$ represents a random phase for each $\alpha$. The amplitude $|\mathbf{A}_{\alpha}|$, according to \cite{chandran2010perpendicular} follows

\begin{align}
        |\mathbf{A}_{\alpha}| =  CB_{0}\left( |\mathbf{k}_{\alpha,\perp}|d_{i} \right)^{-1/3}
\end{align}

\noindent Thus, the components of the total initial magnetic variations are 

\begin{align}
    \delta B_{T,x} = D\sum_{\alpha=1}^{8}\delta B_{\alpha,x} \quad
    \text{and} \quad \delta B_{T,y} = D\sum_{\alpha=1}^{8}\delta B_{\alpha,y}
\end{align}

\noindent where $D$ is a normalization constant defined as 

\begin{align}
    D=\frac{B_{0}}{ \sqrt{\sum_{\alpha=1}^{8} |\mathbf{A}_{\alpha}|^{2}}}
\end{align}

\noindent which ensures that the total amplitude of all modes $|\delta \mathbf{B}_{T} / B_{0}| \sim 1$ at the beginning of the simulation. We assume that the nonlinear time is comparable to the linear time at the initial time, thus we initialise the simulation with strong turbulence. The nonlinearity parameter $\chi=(\delta B_{T} /B_{0})/(k_{\parallel}/k_{\perp}) \sim L_{Z}/L_{x} \sim 5.2$ at the initial time which quantitatively states that the initialised turbulence is strong. The components of the velocity fluctuations $ \delta \mathbf{u}_{T}$ are calculated self-consistently according to Eq. (\ref{eqn:alfpol}).

\indent The wavelengths of the initial waves at $k_{\perp}d_{i}=1$, are  $\lambda_{\perp} = 2\pi d_{i}$ and $\lambda_{\parallel} = 2\pi/10^{-4/3} d_{i}$. Therefore, the size of the box required to simulate our initial ($m=1$) anisotropic Alfvén waves is $L_{z}=\lambda_{\parallel}$  and $L_{x}=L_{y}=\sqrt{2}\lambda_{\perp}$. However, we use  $L_{z}=125d_{i}$, $L_{x}=L_{y}=24d_{i}$, $\lambda_{\parallel} = L_z$ and $\lambda_{\perp} = \sqrt{2}L_{x}/4$. This choice keeps the ratio $\lambda_{\perp}/\lambda_{\parallel}\approx 10^{-4/3}$ while allowing a wider spatial evolution in the perpendicular direction.

\indent The critical-balance scaling $k_{\parallel} \sim k_{\perp}^{2/3}$ applies to Alfvén waves in the inertial range. The initial fluctuations in our simulation have $k_\perp d_{i} \sim 1$ which is at the transition scale from the inertial to the dissipation range. Natural fluctuations at this scale have an anisotropy consistent with the critical-balance scaling based on the size of the inertial range \citep{wicks2010power}. The scale dependence of the anisotropy in the inertial range also varies when considering dynamic alignment and intermittency \citep{cho2004anisotropy,boldyrev2011spectral, chandran2015intermittency, chen2016recent}. We assume a critical-balance scaling over an inertial range of four decades to capture the relative amplitude of the anisotropy without including the true evolution of the inertial-range turbulence. Therefore we initialise with fluctuations at $k_{\perp}d_{i} \sim 1$ that have such an anisotropy. The wavevector anisotropy in the dissipation range is less well understood and at kinetic scales it is not clear whether the turbulence is mostly carried by KAWs, whistler waves or a combination of compressive and non-compressive modes \citep{schekochihin2009astrophysical, chen2010interpreting,  boldyrev2012spectrum}. Moreover, pressured-balanced structures also contribute to the turbulent cascade \citep{verscharen2012kinetic, narita2015kinetic, verscharen2017kinetic}. Nevertheless, our anisotropic initialisation is supported by solar-wind measurements \citep{horbury2008anisotropic, alexandrova2009universality, wicks2010power, wicks2011anisotropy} and allows a kinetic cascade to develop self-consistently as the simulation evolves.


\section{Second-order structure function}
\label{app:sosf}
Following \citet{cho2000anisotropy}, we define the local magnetic field between two points $\mathbf{r}_{1}$ and $\mathbf{r}_{2}$ as 

\begin{align}
    \mathbf{B}_{l}=\frac{\mathbf{B}(\mathbf{r}_{2}) + \mathbf{B}(\mathbf{r}_{1}) }{2}.
\end{align}

\begin{figure}
\centering
\begin{subfigure}[b]{0.49\linewidth}
\includegraphics[width=\textwidth]{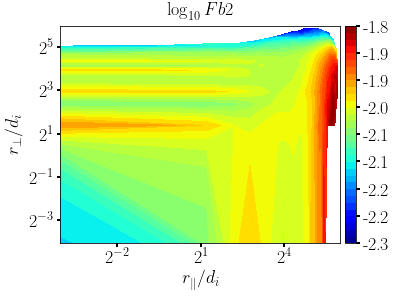}
\caption{$t=0$}
\label{fig:sosf0}
\end{subfigure}
\begin{subfigure}[b]{0.49\linewidth}
\includegraphics[width=\textwidth]{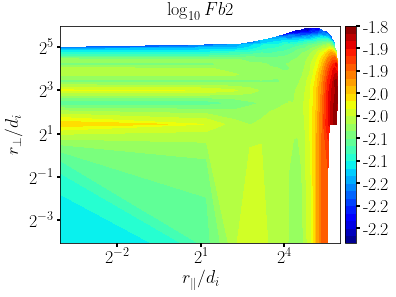}
\caption{$t=12/\omega_{pi}$}
\label{fig:sosf12}
\end{subfigure}
\begin{subfigure}[b]{0.49\linewidth}
\includegraphics[width=\textwidth]{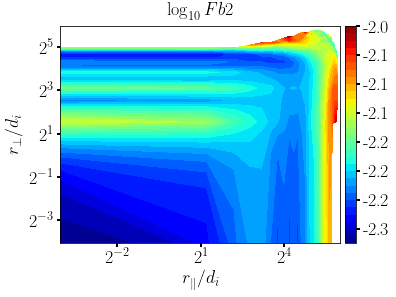}
\caption{$t=120/\omega_{pi}$}
\label{fig:sosf120}
\end{subfigure}
\begin{subfigure}[b]{0.49\linewidth}
\includegraphics[width=\textwidth]{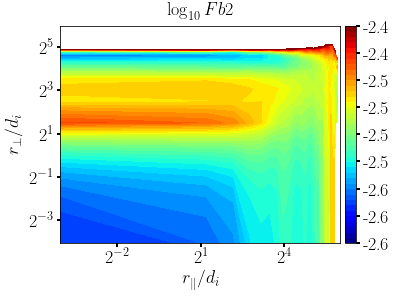}
\caption{$t=240/\omega_{pi}$}
\label{fig:sosf240}
\end{subfigure}
\caption{Second-order structure function of the magnetic fluctuation $\mathbf{b}$ in the $r_{\perp},r_{\parallel}$-plane. $\log_{2} F2b$ at $t=0/\omega_{pi}$ (\textit{a}), $t=12/\omega_{pi}$ (\textit{b}), $t=t_{R}$ (\textit{c}) and $t=240/\omega_{pi}$ (\textit{d}). At $t=0$, while the magnetic energy is distributed across multiple perpendicular scales, it is mainly stored at large parallel scales. At $t=t_{R}$ the magnetic energy is distributed across multiple parallel scales.}  
\label{fig:R_sosf}
\end{figure}

We define the coordinate parallel to $\mathbf{B}_{l}$ as $r_{\parallel}=\hat{z}\cdot (\mathbf{r}_{2} - \mathbf{r}_{1})$ and the coordinate perpendicular as $r_{\perp}=|\hat{z} \times (\mathbf{r}_{2} - \mathbf{r}_{1})|$, where $\hat{z}=\mathbf{B}_{l}/|\mathbf{B}_{l}|$. With these definitions, we calculate the second-order structure function of the magnetic fluctuations $\mathbf{b}(\mathbf{r_{1}}) = \mathbf{B}_{l} - \mathbf{B}(\mathbf{r}_{1})$ as 

\begin{align}
    Fb2(r_{\perp},r_{\parallel})= \langle |\mathbf{b}(\mathbf{r}_{2}) -\mathbf{b}(\mathbf{r}_{1})|^{2}\rangle,
\end{align}

\noindent where $\langle \ \rangle$ represents the average over the spatial domain. In order to discretize the $r_{\perp}r_{\parallel}$-plane, we calculate the values of $r_{\perp}$, $r_{\parallel}$ and $Fb2$ for each pair of points $\mathbf{r}_{1},\mathbf{r}_{2}$. Then, for each pixel, we calculate the mean value as the sum of all $Fb2$ divided by the number of combinations ($\mathbf{r}_1$,$\mathbf{r}_2$) in each pixel. We apply a filter to remove the pixels with less than $\sqrt N$ combinations, where $N$ is the total number of combinations in the $r_{\perp}$, $r_{\parallel}$ space. \\
Figure \ref{fig:R_sosf} shows $\log(Fb2)$ in the $r_{\perp},r_{\parallel}$-plane for the time steps $t=0, t=12/\omega_{pi}, t=120/\omega_{pi}$ and $t=240/\omega_{pi}$. At $t=12/\omega_{pi}$, the structure function indicates a perpendicular cascade of the magnetic energy. On the other hand, the structure function does not give evidence of a strong parallel cascade and is, instead, still consistent with our initial conditions in terms of the parallel extent of the magnetic-field fluctuations. At $t=120/\omega_{pi}$, the green horizontal structure suggests that the magnetic energy has been redistributed and cascaded to smaller parallel scales. The analysis of the structure functions is consistent with our analysis of the Fourier spectra in Figure \ref{fig:2D12}.        

\bibliographystyle{jpp}
\bibliography{Three_D_MR_PIC_APT}
\end{document}